\batchmode
\makeatletter
\def\input@path{{C:/Users/ayc16/Dropbox/t-hurdles/2024-04-MS-Accept/lyx-thurdles-2024-04/}}
\makeatother
\documentclass[12pt,english]{article}
\usepackage[T1]{fontenc}
\usepackage[latin9]{inputenc}
\usepackage{verbatim}
\usepackage{url}
\usepackage{amsmath}
\usepackage{amsthm}
\usepackage{amssymb}
\usepackage{setspace}
\usepackage[authoryear]{natbib}
\makeatletter
\theoremstyle{plain}
\newtheorem{prop}{\protect\propositionname}
\theoremstyle{plain}
\newtheorem{cor}{\protect\corollaryname}

\usepackage{chenpaper}
\usepackage{chencommands}

\makeatother

\usepackage{babel}
\providecommand{\corollaryname}{Corollary}
\providecommand{\propositionname}{Proposition}

\begin{document}
\title{\textbf{Do \emph{t}-Statistic Hurdles Need to be Raised?}}
\author{\large {Andrew Y. Chen}\\{\normalsize Federal Reserve Board}}
\date{April 2024\thanks{First posted to SSRN: September 25, 2018. I thank Antonio Gil de Rubio,
Preston Harry, Jack McCoy, and Nelson Rayl for excellent research
assistance, Rebecca Wasyk for excellent scientific programming, and
Dino Palazzo and Fabian Winkler for many valuable discussions. I also
thank three anonymous referees, an anonymous associate editor, Christine
Dobridge, Bjorn Eraker, Cam Harvey, Laura Liu, Alan Moreira (discussant),
Nelson Rayl, Alessio Saretto (discussant), Ivan Shaliastovich, Mihail
Velikov, and seminar participants at the Federal Reserve Board, George
Mason University, University of Cologne, and the University of Wisconsin
for helpful comments. The views expressed herein are those of the
authors and do not necessarily reflect the position of the Board of
Governors of the Federal Reserve or the Federal Reserve System.}}
\maketitle
\begin{abstract}
\begin{singlespace}
\noindent Many scholars have called for raising statistical hurdles
to guard against false discoveries in academic publications. I show
these calls may be difficult to justify empirically. Published data
exhibit bias: results that fail to meet existing hurdles are often
unobserved. These unobserved results must be extrapolated, which can
lead to weak identification of revised hurdles. In contrast, statistics
that can target only published findings (e.g. empirical Bayes shrinkage
and the FDR) can be strongly identified, as data on published findings
is plentiful. I demonstrate these results theoretically and in an
empirical analysis of the cross-sectional return predictability literature.
\end{singlespace}
\end{abstract}
\vspace{6ex}

 \thispagestyle{empty}\setcounter{page}{0}

\vspace{5ex}

\pagebreak{}

\section{Introduction}

\setcounter{page}{1} %
\begin{comment}
Motivation
\end{comment}
Suppose a researcher proposes a ``factor'' behind a phenomenon.
How do we determine if this factor is worth noting? At least since
\citet{fisher1925statistical}, researchers have used the following
procedure: (1) construct a statistic that is Student's t-distributed
in the case that the factor is false, and (2) declare the factor a
discovery if this t-statistic exceeds 1.96 in absolute value.\footnote{In the traditional language, a ``false factor'' is called a ``true
null hypothesis'' while a ``true factor'' is called a ``false
null hypothesis.'' Some readers may find the traditional language
confusing, hence my choice of terminology.} More recently, several papers have called for raising this t-statistic
hurdle, or ``t-hurdle'' to guard against false discoveries (\citet*{harvey2016and};
\citet{chordia2020anomalies}), including a paper with dozens of co-authors
(\citet{benjamin2018redefine}).

In this paper, I examine whether these calls can be empirically justified.
An empirical justification is critical, as prior beliefs regarding
cutting edge research are sure to vary across scholars.\footnote{For contrasting prior beliefs regarding asset pricing, see \citet{cochrane2017macro}
and \citet{barberis2018psychology}.} Moreover, an empirical justification seems possible, as multiple
testing statistics provide methods for estimating hurdles that control
the false discovery rate (FDR) (\citet{benjamini1995controlling}).
Indeed, FDR methods are central to \citet*{harvey2016and}'s argument
and the concept is also used in \citet{benjamin2018redefine}.

\begin{comment}
Results summary
\end{comment}

I find that raising the t-hurdle may be difficult to justify empirically.
My results differ from previous studies because I acknowledge weak
identification: the problem that likelihood functions may depend little
on certain model parameters (\citet{canova2009back}). This problem
is important because the data on academic discoveries exhibit publication
bias: results that fail to meet the existing t-hurdle are often unobserved.
Unobserved results need to be extrapolated, which can lead to weak
identification of the key determinants of t-hurdles. I characterize
this problem in a theoretical analysis that extends \citet{benjamini1995controlling}
to a setting with publication bias (as in \citet{hedges1992modeling}).
In an empirical analysis, I bootstrap t-hurdle estimates using a rich
dataset of published cross-sectional stock return predictors (\citet{ChenZimmermann2021}).
Consistent with the theory, the empirical estimates say little about
whether hurdles should be raised, stay the same, or even be lowered.

At the same time, I find other multiple testing statistics are more
strongly identified. Empirical Bayes shrinkage and the FDR among published
factors (\citet{chen2020publication,chen2021most}) focus on the right
tail of t-stats, and this portion of the distribution tends to be
well-observed, in spite of publication bias. For cross-sectional predictors,
I find that published t-stats are biased upward by at most 28\% and
that the FDR for published predictors is at most 22\%, with 95\% confidence.
This strong identification, as well as the weak identification of
t-hurdles, is also found across 10 alternative model specifications.
Robustness is intuitive, as the theoretical results impose few functional
form assumptions. Readers may differ on whether a worst-case FDR of
22\% is satisfactory, but at least some will interpret these results
as implying that $t$-hurdles need not be raised for the cross-sectional
predictability literature.

\begin{figure}[h!]
\caption{\textbf{Why t-hurdles may not need to be raised. }Panels illustrate
two possible scenarios. Bars are stacked. Panel (a) shows a scenario
with 1,000 factors, all of which are false. Panel (b) shows 1,000
false factors and 1,000 true factors, with the distribution of true
factors selected to match data on published t-statistics in \citet*{mccrary2016conservative}.}
\label{fig:intro} \centering

\subfloat[All False Scenario]{\includegraphics[width=0.45\textwidth]{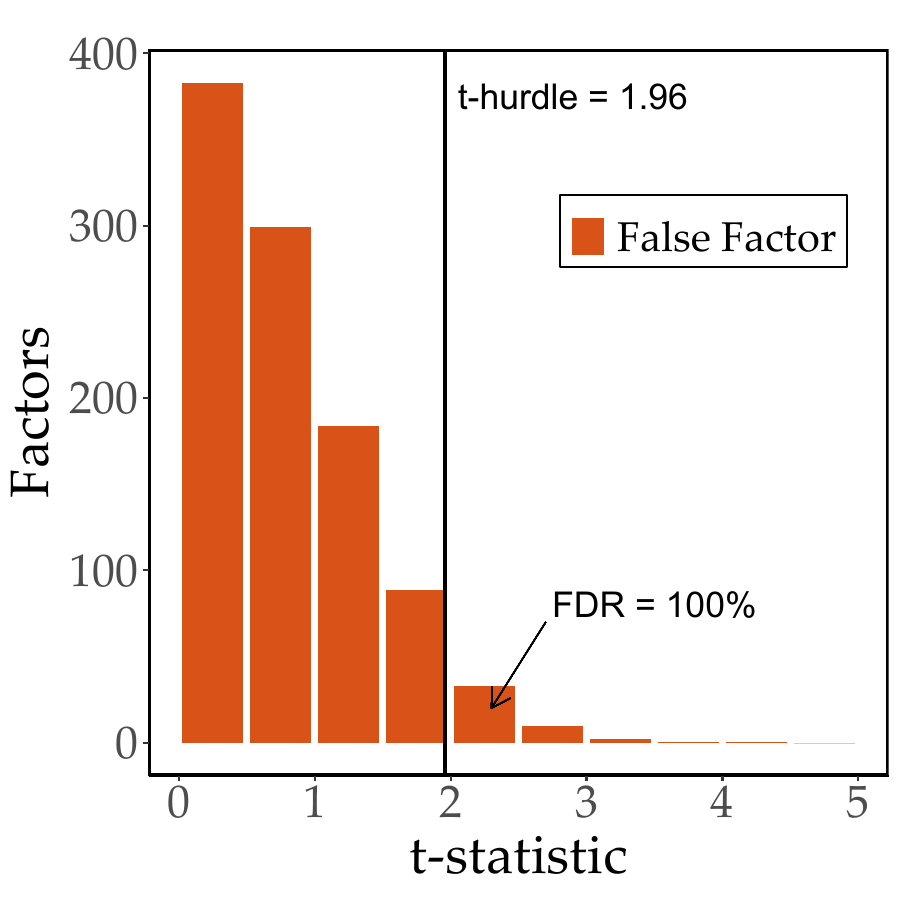}}       
\subfloat[Data-Like Scenario]{\includegraphics[width=0.45\textwidth]{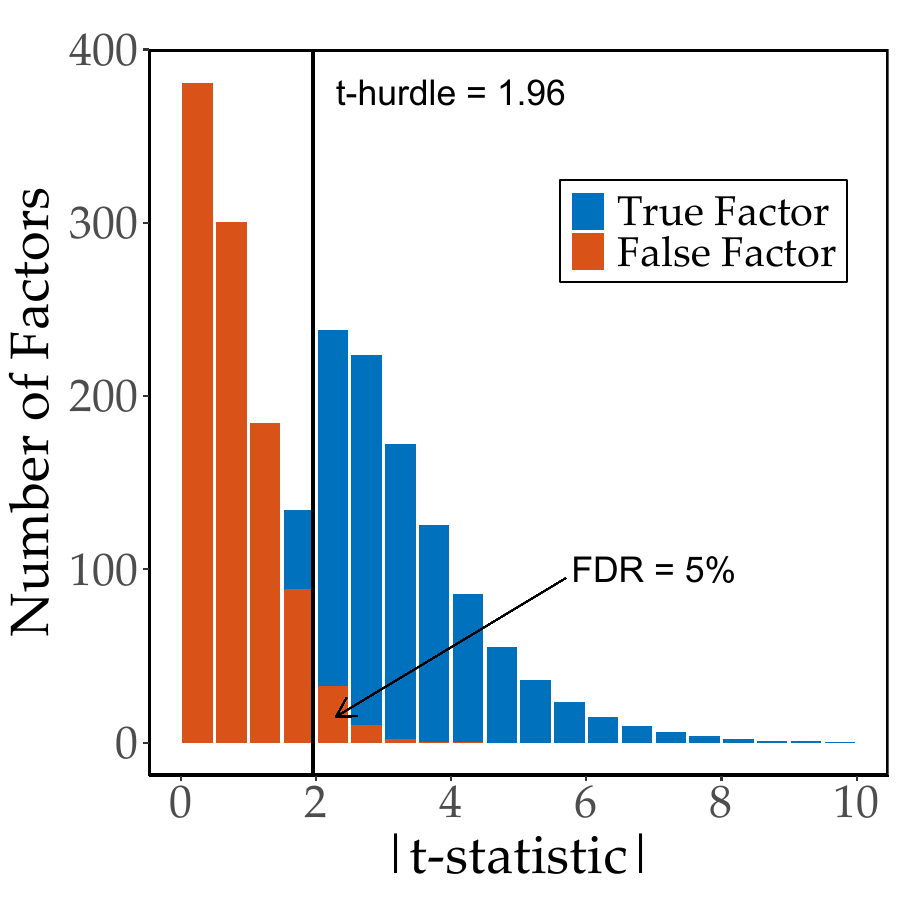}}
\end{figure}

\begin{comment}
Help reader with counterintuitive result and give fun MT intuition
\end{comment}
How can t-hurdles \emph{not }be raised? This idea seems to fly in
the face of multiple testing logic. If one tests 1,000 factors, 5\%
will meet the 1.96 hurdle on average, even if all 1,000 factors are
false. This scenario is visualized in Panel (a) of Figure \ref{fig:intro},
which depicts the distribution of absolute t-stats implied by a standard
normal. By luck, 50 of the 1,000 false factors meet the classical
hurdle. But since all factors are false, the FDR is 50/50 = 100\%.\footnote{Here, the FDR is the number of false and significant factors, divided
by the number of significant factors, in expectation (\citet{benjamini1995controlling}).} Clearly, this scenario implies that the classical hurdle needs to
be raised.

The problem with this logic is that it ignores the information obtained
from multiple testing. Each additional test is another data point,
which brings news about the veracity of factors overall. This news
can in general be positive or negative. More data does not necessarily
mean more bad news.

The data on published factors looks more like Panel (b).\footnote{\citet*{mccrary2016conservative} shows the distribution of t-stats
from meta-studies in political science, psychology, economics, and
all fields of science. Of t-stats that exceed 1.96, roughly half also
exceed 3.0. In Figure \ref{fig:intro}, Panel (b), parameters are
selected to match this moment. A similar distribution is seen in cross-sectional
asset pricing (\citet*{chen2020publication}).} In this panel, there are once again 1,000 false factors (red), so
50 false factors sneak past the 1.96 hurdle. But now there are an
additional 1,000 true factors (blue), and among all factors, 1,000
have t-stats that exceed 1.96. Thus, the 50 false factors represent
only 5\% of the 1,000 factors that meet the classical hurdle. Supposing
that an FDR of 5\% is sufficient, as it is in a variety of fields
from genomics to functional imaging (\citet{benjamini2020selective}),
the classical hurdle need not be raised at all.

Panel (b) illustrates how multiple testing provides more information.
With a single test, asking the question ``what share of t-stats exceed
1.96?'' is impossible to answer. But with multiple testing, answering
this question is a straightforward counting exercise. With many tests
one can also estimate the share of factors that are false (e.g. \citet{storey2002direct}).
If this share is small, then a t-stat of 1.5 may be due to \emph{bad
}luck, in which case the classical hurdle can be \emph{lowered }(\citet{benjamini2000adaptive})\emph{.}

\begin{comment}
Detailed theoretical results:
\end{comment}

My theoretical results characterize exactly when FDR methods imply
a raising of statistical hurdles. The theory revolves around $\pi_{F}$,
the share of false factors among all factors under consideration.
Assuming that an FDR of 5\% is sufficient, the t-hurdle should be
raised if and only if $\pi_{F}$ exceeds the share of t-stats larger
than 1.96. Panel (b) of Figure \ref{fig:intro} shows the knife edge
case in which both of these objects are equal to 50\%, and thus the
classical t-hurdle is sufficient.

The key to empirically justifying a higher t-hurdle, then, is to empirically
justify a large $\pi_{F}$. Unfortunately, my theoretical results
suggest that $\pi_{F}$ is weakly identified under publication bias.
I show that, if false and true factors are in a sense distinct, then
values of $\pi_{F}$ ranging from 0 to $2/3$ imply almost identical
distributions in the region in which data is well-observed. As a result,
empirical studies may say little about the proper t-hurdle.

\begin{comment}
Empirical results
\end{comment}

To quantify this problem, I specify a parametric model of biased publication
and fit it to the \citet{ChenZimmermann2021} dataset using quasi-maximum
likelihood. The model includes the key parameter $\pi_{F}$, as well
as additional parameters that fully describe the distribution of t-stats.
The estimated parameters, in turn, provide consistent formulas for
revised t-hurdles. Bootstrapped estimates show that $\pi_{F}$ is
weakly identified: I re-sample the data, re-estimate the model, and
find that the estimated $\pi_{F}$ ranges from 0\% to 70\% (90\% confidence
interval). t-hurdles that control the FDR at 5\% range from 0 to 3.0
(90\% C.I.).

\begin{comment}
Strong Identification
\end{comment}

More positively, I find that multiple testing statistics that target
published findings are strongly identified, even in the presence of
publication bias. The shrinkage and FDR for published t-stats tend
to be determined by the properties of true factors. And under the
same conditions that imply weak identification of $\pi_{F}$, the
properties of true factors are strongly identified. Intuitively, if
$\pi_{F}$ does not affect the observed distribution, it must be the
properties of true factors that determine the data. Empirical estimates
confirm strong identification for the Chen-Zimmermann dataset. These
results are robust to a cluster-bootstrap that closely mimics correlations
in the empirical data.

An alternative method for handling publication bias is to algorithmically
generate a complete set of factors (\citet{yan2017fundamental,chordia2020anomalies}).
Standard FDR methods can then be applied and $\pi_{F}$ estimated
without the identification problems introduced by publication bias.
This $\pi_{F}$, however, should be interpreted as an upper bound
on the $\pi_{F}$ of the literature, as expert researchers should
be able to find true factors at a higher rate than a simple algorithmic
procedure (\citet{chen2021most}).

\begin{comment}
Robustness
\end{comment}

Code to replicate all figures and tables can be found at \url{https://github.com/chenandrewy/qml-pub-bias}.

\subsection{Related Literature}

Many papers examine multiple testing effects in cross-sectional asset
pricing.\footnote{These papers include \citet{harvey2016and,yan2017fundamental,chordia2020anomalies,chen2020publication,jacobs2020anomalies,chen2021limits,chen2021most,harvey2021uncovering};
and \citet{jensen2021there}.} Among these papers, \citet{harvey2016and}; \citet{chordia2020anomalies};
and \citet{harvey2021uncovering} focus on t-hurdle corrections. To
these papers, I add a framework for understanding when t-hurdles need
to be raised. The framework shows that most estimates in \citet{harvey2016and}
and \citet{chordia2020anomalies} \emph{assume }that t-stat hurdles
need to be raised, and thus they cannot answer the question of whether
t-hurdles need to be raised. It also highlights how the multiple testing
algorithm perhaps most common in finance (\citet{benjamini2001control}
Theorem 1.3) is likely too conservative.

I also add to \citet{harvey2016and}; \citet{chordia2020anomalies};
and \citet{harvey2021uncovering} by examining the identification
problems that come with publication bias (\citet{copas1999works,hedges2005selection}).
These problems would be seen in standard errors on t-hurdles, which
are not provided by \citet{harvey2016and} or \citet{harvey2021uncovering}.
I revisit their estimates and find that the standard errors are so
wide they say little about whether t-stat hurdles should be raised
or even be lowered.

In a contemporaneous paper, \citet{harvey2021uncovering} critique
three assumptions that have been used in this literature: (1) cross-factor
correlations are all equal, (2) the publication probability is a strict
t-stat cutoff, and (3) the selected parametric models may be too restrictive.
I find that none of these issues has a significant effect on the small
shrinkage and FDR estimates found by \citet{chen2020publication}
and \citet{jensen2021there}. I find similar estimates assuming (1)
weak dependence across factors, (2) a smoothly increasing publication
probability, and (3) many distinct parametric forms for latent effect
sizes. Harvey and Liu propose instead to model latent effect sizes
by randomly drawing from empirical data on published and data-mined
factors. This method deviates strongly from the literature (\citet{efron2012large,andrews2019identification})
and it has not been shown to recover effect sizes either theoretically
or in simulations.

Identification under publication bias is also studied in \citet{andrews2019identification},
who prove identification of a non-parametric model. Their proof obtains
the latent distributions by solving a system of ODEs, effectively
assuming an unlimited sample of published tests. My analysis is less
restrictive about data availability. Also unlike Andrews and Kasy,
I connect to the literature on false discovery rates and empirical
Bayes shrinkage. More broadly, my paper helps bridge the literatures
on publication bias (\citet{hedges2005selection,mcshane2016adjusting})
and multiple testing (\citet{efron2012large}) and provides guidance
on which multiple testing corrections should be applied under publication
bias.

There are other arguments against raising t-hurdles. Raising t-hurdles
would lead to published data that is in a sense more distorted. This
reasoning motivates, in part, the push for pre-analysis plans, which
lower t-hurdles provided that the analysis is rigorously pre-specified
(\citet{olken2015promises,kasy2023optimal}). Raising t-hurdles also
fails to address common misinterpretations of statistical significance
and artificial dichotimization introduced by hypothesis testing (\citet{mcshane2019abandon,chen2023publication}).

\section{Theoretical Results\label{sec:theory}}

I describe a general model (Section \ref{sec:theory-model}), define
multiple testing statistics (Section \ref{sec:theory-hurdle-def}),
and then prove results that illustrate weak and strong identification
(Sections \ref{sec:theory-hurdles}-\ref{sec:theory-strong}). Section
\ref{sec:theory-conservative} explains why some commonly-used FDR
methods cannot answer the question of whether t-stat hurdles should
be raised. Section \ref{sec:theory-fnr} discusses false negative
rates.

\subsection{A Model of Multiple Testing and Publication Bias \label{sec:theory-model}}

A literature is generated in two steps. In the first step, researchers
generate an ``unbiased'' set of $N$ factors that obey the assumptions
in \citet{benjamini1995controlling}. Factor $i$ has a t-stat $t_{i}$
that follows 
\begin{align}
t_{i}|\mu_{i} & \sim\text{Normal}\left(\mu_{i},1\right)\label{eq:theory-t_mu}
\end{align}
where $\mu_{i}$ is the latent effect size of factor $i$ (e.g. the
expected return) divided by its standard error. $\mu_{i}$ can also
be thought of as the ``corrected'' t-stat, since it corrects $t_{i}$
for sampling error. $\mu_{i}$ depends on whether factor $i$ is false
($F_{i}$) or true ($T_{i}$)
\begin{align}
\mu_{i}|F_{i} & =0.\label{eq:theory-mu_F}\\
\mu_{i}|T_{i} & \sim g\left(\cdot|\lambda\right),\label{eq:theory-mu_T}
\end{align}
where $g\left(\cdot|\lambda\right)$ is an arbitrary probability distribution
and $\lambda$ is a vector of parameters. Last, and most importantly,
factor $i$ is false with probability $\pi_{F}$.\footnote{\citet{benjamini1995controlling} leave open the possibility of other
distributional assumptions for Equation (\ref{eq:theory-t_mu}), but
the standard normal is commonly used in practice (\citet{efron2012large};
\citet{harvey2016and}). \citet{chen2021limits} shows that the standard
normal assumption holds for long-short portfolios from the asset pricing
literature.}

I describe the factors from the first step as ``unbiased'' because
$t_{i}|F_{i}\sim\text{Normal}\left(0,1\right),$ and thus the theory
of \citet{fisher1925statistical} and \citet{benjamini1995controlling}
applies. However, the second step of literature generation creates
a bias.

In the second step, factor $i$ is published depending on its statistical
significance:
\begin{align}
\Pr\left(\pub_{i}\left||t_{i}|\right.\right) & =s\left(|t_{i}|\right)\label{eq:theory-pubprob}\\
s\left(|t_{i}|\right) & \quad\text{weakly increasing}\label{eq:theory-pubprob-increase}\\
s\left(|t_{i}|\right) & =\bar{s}\quad\text{for \ensuremath{|t_{i}|>\tgood}}\label{eq:theory-pubprob-max}
\end{align}
where $\pub_{i}$ is the event that factor $i$ is published, $s\left(|t_{i}|\right)$
is a function with values in $[0,1]$, and $\bar{s}$ and $\tgood$
are constants. Conceptually, $s(|t_{i}|)$ captures the prevailing
t-stat hurdles that were applied in the generation of the published
data.

This second step implies $t_{i}|\left(F_{i},\pub_{i}\right)\nsim\text{Normal}\left(0,1\right)$,
violating the assumptions of both \citet{fisher1925statistical} and
\citet{benjamini1995controlling}. Thus, to estimate the FDR and related
statistics, one needs to recover the properties of the unbiased factors
generated in the first step.

Equations (\ref{eq:theory-pubprob})-(\ref{eq:theory-pubprob-max})
describe a ``satisficing'' model of publication bias. They say that
a higher t-stat implies a higher probability of publication but once
the t-stat is high enough the community is satisfied and there are
no additional gains. A satisficing model is consistent with the view
that researchers are primarily incentivized to uncover convincing
and interesting mechanisms and that statistical significance plays
a secondary role. This assumption nests the functional forms used
in previous estimates of publication bias (\citet{harvey2016and,andrews2019identification,chen2020publication}).

I refer to t-stats that exceed $\tgood$ as ``well-observed.'' $\tgood$
is typically 1.96 (\citet{andrews2019identification}) or a number
between 1.96 and 3.0 (\citet{harvey2016and,chen2020publication}).
Relatively few t-stats below $\tgood$ are observed and the data moments
in this region are distorted. In contrast, the likelihood conditional
on $|t_{i}|>\tgood$ is not distorted, so one can make inferences
about $\pi_{F}$ and $\lambda$ directly from the conditional moments
in this region.

In reality, the probability of publication depends not only on $|t_{i}|$
but also on supporting evidence. However, data on supporting evidence
is rarely available, so the publication bias literature focuses on
Equation (\ref{eq:theory-pubprob}), which ignores supporting evidence.
The net effect of this misspecification is unclear. Omitting supporting
evidence tends to bias estimates of $\pi_{F}$ and $\lambda$ in a
way that implies stronger latent effects, as these parameters must
absorb the larger $|t_{i}|$ values induced by the supporting evidence.
On the other hand, omitting supporting evidence tends to imply weaker
latent effects, as the supporting evidence is by definition a signal
of strong latent effects. Appendix \ref{sec:app-zevidence} formalizes
these issues and provides some simulation evidence suggesting that
the net bias is small.

The model abstracts from dynamic issues like out-of-sample decay in
effect size. More generally, the effect size depends on whether it
is measured in the original sample ($\mu_{i}$) or out-of-sample ($\mu_{i}^{\text{OOS}}$).
While researchers typically aim to find factors with stable effect
sizes, empirical evidence in cross-sectional asset pricing finds $\mu_{i}^{\text{OOS}}\approx0.50\mu_{i}$
(\citet{mclean2016does,chen2020publication}). Thus, finding that
$\mu_{i}$ is close to $t_{i}$ does not necessarily imply out-of-sample
robustness.

\subsection{Hurdles that Control the False Discovery Rate\label{sec:theory-hurdle-def}}

Calls for raising statistical hurdles often come down to controlling
the false discovery rate with a t-stat hurdle:
\begin{align}
\text{hurdle}\left(5\%\right) & \equiv\min_{h\in\mathbb{R}_{+}}\left\{ h:\Fdr\left(h\right)\le5\%\right\} .\label{eq:theory-hurdle}
\end{align}
where $\Fdr\left(h\right)$ is the Bayesian formulation from \citet{efron2008microarrays}:
\begin{align}
\Fdr\left(h\right) & \equiv\Pr\left(F_{i}\left||t_{i}|>h\right.\right).\label{eq:theory-Fdr-Bayes}
\end{align}
(see also \citet{efron2001empirical} and \citet{storey2002direct}).
Equation (\ref{eq:theory-hurdle}) finds the lowest t-stat hurdle
that results in less than 5\% of factors being false, where 5\% is
chosen for ease of exposition (the main results hold for other choices).
$\Fdr\left(h\right)$ is used in both \citet{ioannidis2005most} and
\citet{benjamin2018redefine}. \citet{jensen2021there} estimates
an object similar to $\Fdr\left(1.96\right)$ for cross-sectional
return predictors. The \citeyearpar{chen2018publication} version
of \citet{chen2020publication} estimates both $\text{hurdle}\left(5\%\right)$
and $\Fdr\left(h\right)$.

As shown by \citet{storey2002direct} and \citet{storey2004strong},
the Bayesian formulation is equivalent to the Benjamini-Hochberg (\citeyear{benjamini1995controlling,benjamini2000adaptive})
approach under weak dependence. To see this, define the Benjamini-Hochberg
FDR:
\begin{align}
\FDR_{BH}\left(h\right) & \equiv\E\left[\frac{\sum_{i=1}^{N}I\left(F_{i}\cap|t_{i}|>h\right)}{\sum_{i=1}^{N}I\left(|t_{i}|>h\right)}\right],\label{eq:theory-FDR-BH}
\end{align}
where $I\left(\cdot\right)$ is an indicator function and I assume
$\sum_{i=1}^{N}I\left(|t_{i}|>h\right)>0$. Then suppose the following
weak law of large numbers hold: For any $h\in\mathbb{R}_{+}$,
\begin{align}
\frac{1}{N}\sum_{i=1}^{N}I\left(|t_{i}|>h\right) & \overset{p}{\rightarrow}\Pr\left(|t_{1}|>h\right)\label{eq:theory-wlln-1}\\
\frac{1}{\pi_{F}N}\sum_{i=1}^{N}I\left(F_{i}\cap|t_{i}|>h\right) & \overset{p}{\rightarrow}\Pr\left(|t_{1}|>h|F_{1}\right).\label{eq:theory-wlln-2}
\end{align}
These expressions say that if you keep counting the shares of factors
that exceed a hurdle, you'll eventually find the probability that
a factor exceeds a hurdle. Rearranging Equation (\ref{eq:theory-FDR-BH})
and plugging in Equations (\ref{eq:theory-wlln-1})-(\ref{eq:theory-wlln-2})
leads to Equation (\ref{eq:theory-Fdr-Bayes}):
\begin{align}
\E\left[\frac{\sum_{i=1}^{N}I\left(F_{i}\cap|t_{i}|>h\right)}{\sum_{i=1}^{N}I\left(|t_{i}|>h\right)}\right] & =\E\left[\frac{\sum_{i=1}^{N}I\left(F_{i}\cap|t_{i}|>h\right)/\left(\pi_{F}N\right)}{\sum_{i=1}^{N}I\left(|t_{i}|>h\right)/N}\frac{\pi_{F}N}{N}\right]\nonumber \\
 & \overset{p}{\rightarrow}\frac{\Pr\left(|t_{i}|>h|F_{i}\right)\Pr\left(F_{i}\right)}{\Pr\left(|t_{i}|>h\right)}=\Pr\left(F_{i}\left||t_{i}|>h\right.\right),\label{eq:theory-FDR-equiv}
\end{align}
where the last equality applies Bayes rule. \citet{harvey2016and}'s
Section 4 uses Equation (\ref{eq:theory-FDR-BH}) in the constraint
of Equation (\ref{eq:theory-hurdle}) to argue that t-stat hurdles
need to be raised.

\subsection{Weak Identification of t-stat Hurdles\label{sec:theory-hurdles}}

As discussed in \citet{benjamini2000adaptive} (Remark 2), Equations
(\ref{eq:theory-hurdle}) and (\ref{eq:theory-Fdr-Bayes}) imply that
hurdles need not be raised and may even be lowered. The following
proposition characterizes exactly when this occurs:
\begin{prop}
\label{prop:main}If $\,\Pr\left(F_{i}\left||t_{i}|>h\right.\right)$
is strictly decreasing in $h$, then 
\begin{align}
 & \text{hurdle}\left(5\%\right)>1.96\text{ if and only if }\text{\ensuremath{\pi_{F}}}>\Pr\left(|t_{i}|>1.96\right)\label{eq:theory-prop-result-1}\\
 & \text{hurdle}\left(5\%\right)=1.96\text{ if and only if }\text{\ensuremath{\pi_{F}}}=\Pr\left(|t_{i}|>1.96\right)\\
 & \text{hurdle}\left(5\%\right)<1.96\text{ if and only if }\text{\ensuremath{\pi_{F}}}<\Pr\left(|t_{i}|>1.96\right)\label{eq:theory-prop-result-3}
\end{align}
\end{prop}
The proof is in Appendix \ref{sec:app-proof-main}.

Proposition \ref{prop:main} says the t-stat hurdle needs to be raised
if and only if $\pi_{F}$ is larger than $\Pr\left(|t_{i}|>1.96\right)$.
This comparison is visualized in Panel (b) of Figure \ref{fig:intro}.
$\pi_{F}$ is the share of false factors (red) relative to the total
mass, while $\Pr\left(|t_{i}|>1.96\right)$ is the mass to the right
of the vertical line. In this illustration, these probabilities are
equal, so the multiple testing hurdle is equal to the classical one.
In contrast, Panel (a) shows a setting in which the share of the red
mass (100\%) is much larger than the mass to the right of the vertical
line (5\%), implying that hurdle needs to be raised.

Some readers may have the intuition that $\pi_{F}>\Pr\left(|t_{i}|>1.96\right)$
must be the ``right'' case of Proposition \ref{prop:main}. This
intuition is perhaps natural, as statistics is a conservative institution.
However, there is no logical reason for why we must have $\pi_{F}>\Pr\left(|t_{i}|>1.96\right)$.
Assuming that this expression holds amounts to placing additional
restrictions on the model of Section \ref{sec:theory-model} and it
is unclear how to motivate these restrictions. Thus, it is the role
of data to tell us which case of Proposition \ref{prop:main} is the
right one. For example, \citet{benjamini2006adaptive} reviews several
methods for estimating $\pi_{F}$.

The methods in \citet{benjamini2006adaptive}, however, assume that
all factors are observed. Under publication bias (Equations (\ref{eq:theory-pubprob})-(\ref{eq:theory-pubprob-max}))
it may be difficult to identify $\pi_{F}$. Indeed, the meta-study
literature on publication bias runs into flat likelihood functions
and estimation problems for other parameters (\citet{copas1999works}
and \citet{hedges2005selection}). The following proposition shows
publication bias leads to identification problems for $\pi_{F}$:
\begin{prop}
\label{prop:weak}(Weak Identification) If there exists $\varepsilon$
such that for all $\bar{t}>\tgood$,
\begin{align}
\frac{\Pr\left(|t_{i}|\in[\tgood,\bar{t}]\bigl|F_{i}\right)}{\Pr\left(|t_{i}|\in[\tgood,\bar{t}]|T_{i}\right)} & <\varepsilon\label{eq:theory-weak-cond}
\end{align}
then for any $\pi\in[0,2/3]$, $\pi'\in[0,2/3]$, and $\bar{t}>\tgood$,
\begin{align}
\left|\frac{\Pr\left(|t_{i}|\le\bar{t}||t_{i}|>\tgood\right)|_{\pi_{F}=\pi'}}{\Pr\left(|t_{i}|\le\bar{t}||t_{i}|>\tgood\right)|_{\pi_{F}=\pi}}-1\right| & <2\varepsilon+O\left(\varepsilon^{2}\right),\label{eq:theory-weak-result}
\end{align}
where $\Pr\left(|t_{i}|\le\bar{t}||t_{i}|>\tgood\right)|_{\pi_{F}=\pi}$
is 
\begin{align}
\Pr\left(|t_{i}|\le\bar{t}||t_{i}|>\tgood\right) & =\frac{\pi_{F}\Pr\left(|t_{i}|\in(\tgood,\bar{t}]\bigl|F_{i}\right)+\left(1-\pi_{F}\right)\Pr\left(|t_{i}|\in(\tgood,\bar{t}]\bigl|T_{i}\right)}{\pi_{F}\Pr\left(|t_{i}|>\tgood|F_{i}\right)+\left(1-\pi_{F}\right)\Pr\left(|t_{i}|>\tgood\bigl|T_{i}\right)}.\label{eq:theory-cond-CDF}
\end{align}
evaluated at $\pi_{F}=\pi$.
\end{prop}
The proof is in Appendix \ref{sec:app-proof-weak}.

Proposition 2 says that, under certain assumptions, $\pi_{F}$ has
little effect on model predictions. In particular, Equation (\ref{eq:theory-weak-result})
says the distribution of t-stats in the well-observed region changes
by at most a factor of $2\varepsilon$ across values of $\pi_{F}$
ranging from 0 to 2/3. The bound of 2/3 is chosen for illustrative
purposes. More generally, a bound of $\bar{\pi}$ for $\pi_{F}$ implies
an upper bound of $\bar{\pi}/(1-\bar{\pi})\varepsilon$ on the RHS
of Equation (\ref{eq:theory-weak-result}) (see proof).

The key assumption is Equation (\ref{eq:theory-weak-cond}), which
formalizes the implicit assumption in hypotheses testing that true
and factors are distinct. If true and false factors are not distinct,
then the pursuit of separating true from false factors is in a sense
misguided. In my setting, false factors by definition have $|t_{i}|$
close to zero, so distinctiveness can be thought of as saying that
true factors are more likely to have $|t_{i}|>\tgood$. Equation (\ref{eq:theory-weak-cond})
says that true factors are $1/\varepsilon$ times more likely to be
found in this region than false factors, where $\varepsilon$ is presumably
small.

Figure \ref{fig:hlz} illustrates Proposition \ref{prop:weak} by
overlaying alternative parameterizations of \citet*{harvey2016and}'s
parametric model. HLZ's baseline estimate implies $\pi_{F}=0.444$
and $g\left(\cdot|\lambda\right)$ is an exponential distribution
with mean of about 2. This estimate implies that raising t-hurdles
is necessary. According to HLZ's Table 5, the classical t-hurdle of
1.96 needs to be raised to 2.27 to control the FDR at 5\%. The distribution
of t-stats implied by HLZ's baseline estimate is shown in the blue
bars of Figure \ref{fig:hlz}.
\begin{center}
{[}Figure \ref{fig:hlz} about here{]}
\par\end{center}

But now consider an alternative model, that uses all the same parameters
as HLZ's baseline but simply changes $\pi_{F}$ to zero. With no false
factors, the alternative model implies that t-hurdles can be lowered,
all the way to 0. The predicted distribution of t-stats is shown in
the red bars of Figure \ref{fig:hlz}.

These two models differ markedly in their densities of t-stats near
zero. However, these t-stats are difficult to observe and must be
extrapolated. Indeed, HLZ assume $\tgood=2.57$ (vertical line), and
for $|t_{i}|>\tgood$ the two distributions are nearly identical.
Thus, it is unlikely that the data can tell us whether t-stats should
be raised, stay the same, or even be lowered.

This identification problem should show up in standard error estimates.
They imply that small perturbations to the data imply very different
t-stat hurdles, and thus large standard errors. HLZ do not provide
standard errors for their hurdle estimates. I fill this gap in Section
\ref{sec:emp}.

\subsection{Strongly-Identified Multiple Testing Statistics\label{sec:theory-strong}}

Revised hurdles are not the only way to deal with a many-factor setting.
Chapter 1 of \citet{efron2012large}'s textbook on large scale inference
examines empirical Bayes shrinkage: 
\begin{align}
\text{shrinkage}\left(\bar{t}\right) & \equiv\frac{|\bar{t}|-E\left(\mu_{i}\left||t_{i}|=\bar{t}\right.\right)}{|\bar{t}|}\label{eq:theory-shrink-def}
\end{align}
where $\bar{t}\in\mathbb{R}_{+}$. Equation (\ref{eq:theory-shrink-def})
measures how much you should shrink an observed t-stat ($|t_{i}|$)
toward zero in order to recover its unbiased counterpart ($E\left(\mu_{i}\left||t_{i}|=\bar{t}\right.\right)$).
For example, $\text{shrinkage}\left(4.0\right)=0.25$ means that a
t-stat of 4.0 should be shrunk by 25\% and that the unbiased t-stat
is $4.0\times(1-0.25)=3.0$. Since the sample mean return is proportional
to $t_{i}$, the same shrinkage adjustment applies to the sample mean
return. Asset pricing papers that study empirical Bayes shrinkage
include \citet{chen2020publication}, \citet{chinco2021estimating},
\citet{chen2023zeroing}, and \citet{jensen2021there}.

Chapter 2 of \citet{efron2012large} examines the local FDR: 
\begin{align}
\fdr\left(\bar{t}\right) & \equiv\Pr\left(F_{i}\left||t_{i}|=\bar{t}\right.\right).\label{eq:theory-localfdr}
\end{align}
$\fdr\left(\bar{t}\right)$ is just the probability that a factor
with a t-stat of $\bar{t}$ is false. Integrating $\fdr\left(\bar{t}\right)$
from $h$ to infinity leads to Equation (\ref{eq:theory-Fdr-Bayes}).

In contrast to $\text{hurdle}\left(5\%\right)$, empirical Bayes shrinkage
and the local fdr can be used to target only published findings (by
selecting $\bar{t}$ to be equal to the t-stats of published factors).
And in the published region, shrinkage and the local fdr depend more
so on the parameter vector that governs true factors ($\lambda$).
To see this, note that the key term in Equation (\ref{eq:theory-shrink-def})
can be written as 
\begin{align}
E\left(\mu_{i}\left||t_{i}|=\bar{t}\right.\right) & =E\left(\mu_{i}\left||t_{i}|=\bar{t},T_{i}\right.\right)\left[\frac{1}{1+\Delta}\right]\label{eq:theory-bias-approx}\\
\Delta & \equiv\left(\frac{\pi_{F}}{1-\pi_{F}}\right)\left[\frac{f_{|t||F}\left(\bar{t}\right)}{f_{|t||T}\left(\bar{t}\right)}\right]\label{eq:theory-bias-approx2}
\end{align}
where $f_{|t||F}\left(\bar{t}\right)$ and $f_{|t||T}\left(\bar{t}\right)$
are the densities of $|t_{i}|$ given that $i$ is false or true,
respectively. Intuitively, shrinkage of the full model is the same
as shrinkage assuming $\pi_{F}=0$, but with a correction that increases
in $\pi_{F}$. A similar argument can be made for the local FDR. Appendix
\ref{sec:app-bias-approx} provides the derivations.

For empirically relevant $\bar{t}$ and $\lambda$, $\Delta$ is often
small. For example, the HLZ dataset has a mean $|t_{i}|$ of around
4.0, corresponding to $f_{|t||F}\left(4.0\right)=0.00027$. Meanwhile,
their estimates imply $f_{|t||T}\left(4.0\right)=0.076$ , as true
factors are much more likely to have large t-stats. So even if $\pi_{F}$
is as large as 0.9, $\Delta$ is only 0.03. Similarly large t-stats
are commonly found in replications of cross-sectional predictability
papers (\citet{ChenZimmermann2021}). More broadly, \citeauthor{brodeur2016star}
find that t-stats of 4.0 are quite common in the main hypotheses tests
reported in top economics journals.

This analysis rests critically on $\lambda$, which must be estimated
with published data and its associated identification problems. Fortunately,
the following proposition shows that $\lambda$ can be strongly identified
\begin{prop}
\label{prop:strong} (Strongly Identified Parameters) If there exists
$\varepsilon$ such that for all $\bar{t}>\tgood$, 
\begin{align}
\frac{\Pr\left(|t_{i}|\in[\tgood,\bar{t}]\bigl|F_{i}\right)}{\Pr\left(|t_{i}|\in[\tgood,\bar{t}]|T_{i}\right)} & <\varepsilon\label{eq:theory-weak-cond-1}
\end{align}
and $\pi_{F}<2/3$, then 
\begin{align}
\left|\frac{P\left(|t_{i}|<\bar{t}||t_{i}|>\tgood\right)}{P\left(|t_{i}|<\bar{t}||t_{i}|>\tgood,T_{i}\right)}-1\right| & \le2\varepsilon+O\left(\varepsilon^{2}\right).\label{eq:theory-strong-result}
\end{align}
\end{prop}
The proof is in Appendix \ref{sec:app-proof-strong}.

Proposition \ref{prop:strong} says that, under the same condition
that lead to weak identification of t-hurdles (Equation (\ref{eq:theory-weak-cond})),
the model's predictions about well-observed data ($P\left(|t_{i}|<\bar{t}||t_{i}|>\tgood\right)$)
are almost the same as the predictions that come from assuming all
factors are true ($P\left(|t_{i}|<\bar{t}||t_{i}|>\tgood,T_{i}\right)$).
As in the Proposition \ref{prop:weak}, this proposition uses the
$\pi_{F}<2/3$ for illustrative purposes. The more general case in
which $\pi_{F}$ is bounded by $\bar{\pi}$ implies an upper bound
of $\bar{\pi}/(1-\bar{\pi})\varepsilon$ on the RHS of Equation (\ref{eq:theory-strong-result})
(see proof).

An implication of Proposition \ref{prop:strong} is that a reasonable
estimate of $\lambda$ could potentially be found by estimating the
model while assuming $\pi_{F}=0$. This result can be seen in Figure
\ref{fig:hlz}. The red bars are essentially the model that one would
estimate assuming $\pi_{F}=0$. Thus, assuming $\pi_{F}=0$ would
lead to roughly the same $\lambda$ as HLZ's baseline estimate, in
which $\pi_{F}=0.44$. Intuitively, if $\pi_{F}$ is not at all responsible
for fitting the t-stats to the right of 1.96, it must be the other
parameters in HLZ's estimate that accomplish this.

Taken with Equation (\ref{eq:theory-bias-approx}), Proposition \ref{prop:strong}
implies that the bias in large t-stats could potentially be identified
with just published data. This result is seen in the small standard
errors in \citet{chen2020publication}'s shrinkage estimates as well
as the robustness of their estimates to alternative modeling assumptions.
Indeed, Chen and Zimmermann's headline bias estimate of 12\% is not
far from \citet{harvey2021uncovering}'s median estimate of 19\%,
even though the two papers use very different modeling assumptions
and estimation methods. Similar estimates are also found in \citet{chinco2021estimating}
and \citet{jensen2021there}, who use yet another set of methods (see
discussion in \citet{chen2023publication}). Section \ref{sec:emp}
provides additional evidence that these bias estimates are strongly
identified for the cross-sectional predictability literature.

The key to strong identification is flexibility. Unlike, $\text{hurdle}(5\%)$,
$\text{shrinkage}\left(\bar{t}\right)$ and $\text{\fdr\ensuremath{\left(\bar{t}\right)} }$allow
one to focus on the portion of the data that is well-observed. If
the data are not informative about $\text{shrinkage}\left(1.0\right)$,
one can instead examine $\text{shrinkage}\left(4.0\right)$. The cost
of this flexibility is that $\text{shrinkage}\left(\bar{t}\right)$
and $\text{\fdr\ensuremath{\left(\bar{t}\right)}}$ do not provide
a precise answer to the question, ``do t-hurdles need to be raised?''
They provide important evidence (e.g. published t-stats are biased
upward by 12\%). But ultimately an additional framework is required
to convert this evidence into a new t-hurdle.

Also unlike $\text{hurdle}(5\%)$, $\text{shrinkage}\left(\bar{t}\right)$
provides economic magnitudes. Thus, it avoids the potentially artificial
dichotomization that can come from hypothesis testing (\citet{mcshane2019abandon}).
This gain, however, comes at additional analytical costs, namely that
$\text{shrinkage}\left(\bar{t}\right)$ requires selecting $g(\cdot|\lambda)$
and estimating $\lambda$.\footnote{Under publication bias, the local FDR can be bounded without estimating
$\lambda$ by using a kind of worst case scenario (\citet{chen2021most}).}

An important caveat of this analysis is that it relies on $\pi_{F}$
being not too close to 1.0. For $\pi_{F}=0.9$ or higher, the $2\varepsilon$
bound becomes a $9\varepsilon$ bound (or higher), and Proposition
\ref{prop:strong} may have little bite. Intuitively, if almost all
factors are false, then even the extreme right tail of the distribution
will be largely determined by false factors. However, we will see
that empirical evidence suggests $\pi_{F}\ge0.9$ is unlikely cross-sectional
return predictors (Section \ref{sec:emp}).

\subsection{Conservative FDR Controls\label{sec:theory-conservative}}

Some FDR methods avoid estimation of $\pi_{F}$ by effectively assuming
$\pi_{F}\ge1.0$. As a result, these conservative methods \emph{assume}
that t-hurdles need to be raised, and thus cannot answer the question
of \emph{whether} t-hurdles need to be raised.

These conservative algorithms include \citet{benjamini2001control}'s
(BY's) Theorem 1.3, which is emphasized in HLZ, and is popular in
the finance literature. As shown by \citet{storey2002direct}, BY's
Theorem 1.3 is equivalent to replacing the constraint in Equation
(\ref{eq:theory-hurdle}) with an estimator:\footnote{This is seen Equation (13) of \citet{storey2002direct}. See also
the Equivalence Theorem of \citet{efron2002empirical}. Strictly speaking,
these papers show this equivalence for the \citet{benjamini1995controlling}
algorithm, however, the BY algorithm simply modifies the \citet{benjamini1995controlling}
algorithm with a constant factor.} 
\begin{align}
h_{BY1.3}\left(5\%\right) & \equiv\min_{h\in\mathbb{R}_{+}}\left\{ h:\widehat{\FDR}\left(h\right)\le5\%\right\} \label{eq:theory-equiv}
\end{align}
where
\begin{align}
\widehat{\FDR}\left(h\right) & \equiv\frac{\Pr\left(|t_{i}|>h\bigl|F_{i}\right)}{\widehat{\Pr}\left(|t_{i}|>h\right)}\left(\sum_{i=1}^{N}\frac{1}{i}\right)\label{eq:prop-equiv-FDRhat}\\
\widehat{\Pr}\left(|t_{i}|>h\right) & \equiv N^{-1}\sum_{i=1}^{N}I\left(|t_{i}|>h\right).\label{eq:prop-equiv-Prhat}
\end{align}
Compare Equation (\ref{eq:prop-equiv-FDRhat}) to the result of applying
Bayes rule to $\Pr\left(F_{i}\left||t_{i}|>h\right.\right)$
\begin{align}
\Pr\left(F_{i}\left||t_{i}|>h\right.\right) & =\frac{\Pr\left(|t_{i}|>h\left|F_{i}\right.\right)}{\Pr\left(|t_{i}|>h\right)}\pi_{F}\label{eq:theory-fdr-bayes}
\end{align}
Thus, $h_{BY1.3}\left(5\%\right)$ effectively assumes that $\pi_{F}=\sum_{i=1}^{N}\frac{1}{i}$.
In HLZ's case, where $N\approx300$, we have $\pi_{F}=6.3$. So this
algorithm implies that, not only is $\pi_{F}$ is large, but it is
\emph{six times }larger than 1.0. Since $\pi_{F}$>1.0 is physically
impossible, it is hard to justify why such a severe penalty is necessary.
Indeed, $h_{BY1.3}\left(5\%\right)$ is described as ``a severe penalty''
and ``not really necessary'' (\citet{efron2012large}). Even the
original \citet{benjamini2001control} paper says this algorithm is
``very often unneeded, and yields too conservative of a procedure.''

An implication of this conservatism is that BY's Theorem 1.3 always
implies that t-hurdles need to be raised. Assuming $\pi_{F}>1$, only
one case of Proposition \ref{prop:main} is possible:
\begin{cor}
\label{prop:bh-no-answer} If $\widehat{\FDR}\left(\cdot\right)$
is strictly decreasing, then $h_{BY1.3}\left(5\%\right)>1.96$.
\end{cor}
The proof is in Appendix \ref{sec:app-proof-no-answer}.

Similarly, the seminal \citet{benjamini1995controlling} (BH95) algorithm
is equivalent to replacing $\left(\sum_{i=1}^{N}\frac{1}{i}\right)$
with 1.0 in Equation (\ref{eq:prop-equiv-FDRhat}). Thus, BH95 is
isomorphic to using $\pi_{F}=1.0$ in Equation (\ref{eq:theory-hurdle}),
and Proposition \ref{prop:main} implies that BH95 raises t-hurdles
in all cases except for the extreme case in which $\sum_{i=1}^{N}I\left(|t_{i}|>1.96\right)=N$.

The benefit of imposing $\pi_{F}=1.0$ is simplicity. The BH95 hurdle
can be computed for an arbitrary dataset without any additional specifications.
The cost, however, is that BH95 discards all information in the data
that can inform us about $\pi_{F}$. So while BH95 is an excellent
first pass for controlling for multiple testing, it cannot tell us
which case of Proposition \ref{prop:main} holds, and cannot tell
us whether t-hurdles must be raised.\footnote{Unlike the finance literature, the statistics literature tends to
favor the BH95 algorithm over BY's Theorem 1.3 (\citet{benjamini2020selective}).
HLZ argue for using BY's Theorem 1.3, as it controls the FDR under
arbitrary dependence. But BH95 controls the FDR under weak dependence
(\citet{storey2004strong}) and simulations suggest FDR control under
arbitrary dependence if the tests in question use z-statistics (\citet{reiner2007fdr}).}

In their first draft, Benjamini and Hochberg (1995) did \emph{not}
assume that t-stat hurdles should be raised. As described in \citet{benjamini2010discovering},
the 1989 draft recommended a graphical method for estimating $\pi_{F}$.
After many years of rejections, the authors shifted to the conservative
formulation that became the seminal BH95 algorithm. The original algorithm,
along with the result that t-stat hurdles could possibly be lowered,
was eventually published in \citet{benjamini2000adaptive}, and many
statisticians went on to propose additional estimators for $\pi_{F}$
(\citet{efron2001empirical,allison2002mixture,storey2002direct,genovese2004stochastic,benjamini2006adaptive},
among others).

\subsection{False Negative Rates\label{sec:theory-fnr}}

A natural question is how publication bias affects identification
of false negative rates (FNRs), also known as the type II error rate.
This question can be analyzed using a Bayesian formulation of the
FNR:
\begin{align}
\Fnr(h) & \equiv\Pr\left(T_{i}\left||t_{i}|\le h\right.\right)\label{eq:theory-fnr}
\end{align}
(see \citet{efron2012large} Chapter 4.3). $\Fnr(h)$ is the probability
a factor is true, given that the factor fails to meet the hurdle $h$.
One can alternatively define the FNR using factor counts as in Equation
(\ref{eq:theory-FDR-BH}) (see \citet{genovese2002operating}) but
these definitions are equivalent under weak dependence.

Applying Bayes rule shows how the FNR runs into identification issues:
\begin{align}
\Fnr(h) & =1-\frac{\Pr\left(|t_{i}|\le h\left|F_{i}\right.\right)}{\Pr\left(|t_{i}|\le h\right)}\pi_{F}.\label{eq:theory-fnr-intuition}
\end{align}
In the presence of publication bias, $\pi_{F}$ may be weakly identified
(Proposition \ref{prop:weak}). Moreover, the very definition of publication
bias (Equations (\ref{eq:theory-pubprob})-(\ref{eq:theory-pubprob-max}))
implies that the data that bears on $\Pr\left(|t_{i}|\le h\right)$
is limited. Intuitively, to estimate a false negative rate one needs
to know the number of insignificant t-stats, and publication bias
means that insignificant t-stats are poorly observed.

An alternative way to address the FNR is to use the FDR control that
comes closest to achieving the constraint in Equation (\ref{eq:theory-hurdle}).
Achieving this constraint would typically involve forming a point
estimate of $\pi_{F}$ (rather than assuming an upper bound), as is
pursued in this paper.

\section{Empirical Estimates of t-Stat Hurdles\label{sec:emp}}

The theoretical results illustrate weak and strong identification,
but a precise description of these issues requires empirical estimates
of sampling uncertainty. This section provides one such estimate by
bootstrapping estimates using a dataset constructed from asset pricing
publications.

\subsection{Data\label{sec:emp_data}}

My data begins with 207 published cross-sectional stock return predictors
from the \citet{ChenZimmermann2021} (CZ) dataset (March 2022 release).
I focus on their original predictor portfolios, which consists of
long-short portfolios constructed following the procedures in the
original studies.

This dataset has several advantages as a setting for studying publication
bias. CZ show that the replicated t-stats closely match the originals,
which rules out coding errors, fraud, or other more nefarious sources
of bias. CZ also show that the pairwise correlations cluster around
zero (see also \citet*{chen2021most,bessembinder2021time}), suggesting
that the weak dependence assumptions underlying standard estimation
methods are valid (\citet{wooldridge1994estimation}), and that modeling
the correlations will add relatively little to estimation efficiency.
Moreover, the monthly returns in this dataset can be used to account
for correlations in my bootstrapped standard errors. \citet{chen2021limits}
shows the distribution of t-stats is quite similar to the distribution
found in HLZ, which eases comparison. Last, this dataset is publicly
available at www.openassetpricing.com.

To simplify the baseline model, I drop predictors with in-sample t-stats
that fall below 1.96, leading to a sample of 183 predictors. The approach
of dropping t-stats $<$ 1.96 is also used in HLZ. Carefully modeling
the 24 predictors with smaller t-stats requires a relatively complicated
publication probability function and is likely to introduce more identification
problems (\citet{copas1999works}). In the robustness section, I include
t-stats < 1.96 and find broadly similar results (Section \ref{sec:robust-cz-raw}).

Since all of CZ's factors are cross-sectional return predictors, I
refer to them as ``predictors'' in what follows.

\subsection{Structural Model and Estimation\label{sec:emp-model}}

The structural model adds three assumptions to the model of Section
\ref{sec:theory-model}. The additional structure can be thought of
as restrictions that help identify the key model objects (e.g. $\pi_{F},$
$\Pr\left(|t_{i}|>1.96\right)$).

The first is the unbiased t-stat $\mu_{i}$ is log-normal for true
factors 
\begin{align}
\mu_{i}|T_{i} & \sim\text{Log-normal}\left(\lambda_{\mu},\lambda_{\sigma}\right).\label{eq:emp-t-true}
\end{align}
where $\lambda_{\mu}$ and $\lambda_{\sigma}$ are mean and standard
deviation of $\left(\log\mu_{i}\right)|T_{i}$, respectively. This
form is arguably the simplest assumption one can make about $\mu_{i}|T_{i}$
while ensuring that true predictors have an expected return that is
distinct from 0. Section \ref{sec:robust-mudist} shows that alternative
distributional assumptions have little effect on the results.

The second assumption is a functional form for the publication probability
(Equation (\ref{eq:theory-pubprob})). I assume a staircase function
\begin{align}
\Pr\left(\pub_{i}\left||t_{i}|\right.\right) & =\begin{cases}
\eta\bar{s} & \hspace{1ex}\text{if }|t_{i}|\in(1.96,2.58]\\
\bar{s} & \hspace{1ex}\text{if }|t_{i}|>2.58.
\end{cases}\label{eq:emp-pub-prob}
\end{align}
where the t-stat cutoffs of 1.96 and 2.58 correspond to the traditional
5\% and 1\% significance cutoffs, respectively, and $\eta$ represents
a ``haircut'' for marginally-significant predictors. One can think
of this function as the simplest intuitive model of selective publication
for marginally-significant predictors. This functional form is also
assumed in HLZ. Section \ref{sec:robust-pubprob} shows that a logistic
form (as in \citet{chen2020publication} and \citet{harvey2021uncovering})
leads to similar results.

The last assumption is that $\eta$ lies within an interval obtained
from intuition 
\begin{align}
\eta & \in[1/3,2/3].\label{eq:emp-haircut-support}
\end{align}
Equation (\ref{eq:emp-haircut-support}) says that at least some marginally-significant
predictors are missing ($\eta\le2/3$), but at least a significant
minority are reported ($\eta\ge1/3$). HLZ uses the more restrictive
assumption that $\eta=1/2$, though they also examine $\eta=1/3$.
Section \ref{sec:robust-pubprob} shows that relaxing this assumption
or restricting this assumption has little effect on the main results.

I also assume $\pi_{F}\in[0.01,0.99]$ for technical reasons. I use
numerical integration to compute the log-likelihoods and multiple
testing statistics, and I find that these methods can become poorly
behaved for $\pi_{F}$ that is extremely close to 0 or 1.0. One can
increase the range of $\pi_{F}$ with more complicated numerical methods
but this complication is unlikely to affect the main results.

I estimate $\theta\equiv\left(\pi_{F},\lambda_{\mu},\lambda_{\sigma}\right)$
with quasi-maximum likelihood (QML). I choose $\hat{\theta}$ to maximize
the mean log marginal likelihood of $|t_{i}|$ conditional on $i$
being published. I do not estimate $\bar{s}$, as it is not identified
(it is cancelled out in the conditional likelihood). Under technical
conditions, this QML estimator is consistent (\citet*{wooldridge1994estimation,liu2020forecasting}).\footnote{The key technical assumption is that the log marginal likelihood of
a single observation satisfies the uniform weak law of large numbers.} Intuitively, the model implies that the mean derivatives of the marginal
likelihoods are zero when evaluated at the true parameters, and this
implication can be used as moment conditions to estimate the model.

I confirm that QML is consistent under pairwise correlations as high
as 0.90 in Appendix \ref{sec:app-sim-est}. Indeed, in simulated samples
of only 200 t-stats, I find QML is essentially unbiased for $\pi_{F}$
and produces standard errors of around 0.10 to 0.15. These simulations
assume no publication bias and demonstrate that the large standard
errors I find for $\pi_{F}$ are not due to QML.

I measure estimation uncertainty using two bootstraps. My baseline
bootstrap is simple and non-parametric: I draw a full set of t-stats
from the empirical data with replacement and re-run QML. This bootstrap
is transparent and also ensures that the bootstrapped data displays
a similar selection bias and distributional properties as the original
dataset. Moreover, cross-sectional predictor correlations are typically
close to zero (\citet{mclean2016does,ChenZimmermann2021}). These
mild correlations suggest that this simple bootstrap will lead to
similar results as one that accounts for correlations.

For robustness, I also examine a semi-parametric bootstrap that carefully
accounts for correlations. In short, this second approach combines
a cluster bootstrap with a parametric bootstrap. The cluster bootstrap
ensures that the samples closely mimic the correlation structure in
the data while the parametric bootstrap ensures that the samples mimic
the distribution of observed t-stats. Details on the semi-parametric
bootstrap are found in Appendix \ref{sec:app-semiparboot}

\subsection{Parameter Estimates and Intuition\label{sec:emp_fit}}

Table \ref{tab:estimates} shows the resulting parameter estimates.
The point estimate finds that essentially all factors are true ($\hat{\pi}_{F}=0.01$),
but the 90\% confidence interval is huge, ranging from 0.01 to about
0.90. Even the 50\% C.I. is quite large, ranging from 0.01 to about
0.70. There results are consistent with Proposition \ref{prop:weak}
and suggest that t-hurdles are weakly identified.
\noindent \begin{center}
{[}Table \ref{tab:estimates} about here{]}
\par\end{center}

Huge uncertainty about $\pi_{f}$ obtains regardless of whether I
use the simple non-parametric bootstrap (shown in the table with no
parentheses) or the semi-parametric cluster bootstrap (shown with
parentheses). For all parameters in Table \ref{tab:estimates}, accounting
for correlations has almost no effect on estimation uncertainty. This
result is intuitive, as the typical predictor correlation is close
to zero (\citet{ChenZimmermann2021}, see also Figure \ref{fig:semipar-cor}
in the Appendix).

In contrast to $\pi_{F}$, the parameters that govern true factors
($\lambda_{\mu}$ and $\lambda_{\sigma}$) are strongly identified.
Table \ref{tab:estimates} converts these parameters into the expected
unbiased t-stat $\E\left(\mu_{i}|T_{i}\right)$ and standard deviation
of the unbiased t-stat $\SD\left(\mu_{i}|T_{i}\right)$ for ease of
interpretation.\footnote{The lognormal distribution implies 
\begin{align}
\E\left(\mu_{i}|T_{i}\right) & =\exp\left(\lambda_{\mu}+\sigma_{\mu}^{2}/2\right)\\
\text{SD}(\SE_{i}) & =\sqrt{\left[\exp(\lambda_{\mu}^{2})-1\right]\exp(2\lambda_{\mu}+\lambda_{\sigma}^{2})}.
\end{align}
} The bootstrapped estimates imply that $\E\left(\mu_{i}|T_{i}\right)$
is between 2.0 and 3.8 while $\SD\left(\mu_{i}|T_{i}\right)$ is between
1.9 and 2.7, with 90\% confidence. These results are consistent with
Proposition \ref{prop:strong}.

Figure \ref{fig:fit} provides the intuition behind these results.
It compares the predictions of the point estimate with an alternative
model that also uses QML, but the alternative model fixes $\pi_{F}$
at 2/3. Panel (a) shows that both models fit data to the right of
$2.58$ very well. Recall that this is the subsample of the data that
is well-observed (Equation (\ref{eq:emp-pub-prob})). Thus, even a
slight perturbation to the observed data can move the point estimate
from $\pi_{F}=0.01$ to $\pi_{F}=2/3$.
\noindent \begin{center}
{[}Figure \ref{fig:fit} about here{]}
\par\end{center}

The models differ in their predictions about t-stats between 1.96
and 2.58, suggesting the data in this region can help identify $\pi_{F}$.
However, many t-stats in this region may be missing, and it is difficult
know a-priori just how many. Both QML estimates restrict uncertainty
by assuming that the fraction missing is between 1/3 and 2/3 (Equation
(\ref{eq:emp-haircut-support})). Nevertheless, this restriction is
unable to identify $\pi_{F}$. Section \ref{sec:robust-pubprob} shows
that even assuming the missing fraction is 0.5 cannot identify $\pi_{F}$

Thus, the key to identification is the distribution of t-stats below
1.96. As seen in Figure \ref{fig:fit}, the two models have very different
predictions for this region. The data tell us very little about which
is the right story, as there are only 24 observations in this region.
Indeed, a close read of \citet{ChenZimmermann2021} shows that even
these 24 observations are poorly defined, as it is unclear if these
observations should even be called ``predictors.''

This ambiguity leads me to drop t-stats below 1.96 in the estimation
(though they are still shown in Figure \ref{fig:fit}). Nevertheless,
Section \ref{sec:emp-robust} shows that taking on this additional
data still implies substantial uncertainty. Intuitively, only 7 t-stats
below 1.5 are observed, and it is very hard to say how representative
these t-stats are. This very small sample problem suggests that adding
the standard errors of predictor mean returns to the estimation, which
in principle can non-parametrically identify the publication probability
(\citet{andrews2019identification}), will still result in weak identification
of $\pi_{F}$. Indeed, previous versions of this paper did include
these standard errors in the estimation and found similar results.\footnote{Code for the previous versions can be found at \url{https://github.com/chenandrewy/t-hurdles}.}

Panel (b) of Figure \ref{fig:fit} illustrates strong identification.
It examines the distribution of true predictors' t-stats $(t_{i}|T_{i})$
in the point estimate and the alternative model. Despite very different
implications about $\pi_{F}$, both models imply similar distributions
of $t_{i}|T_{i}$. Intuitively, a highly dispersed $t_{i}|T_{i}$
is required to fit the long right tail in observed t-stats, regardless
of $\pi_{F}$. Since $t_{i}|T_{i}$ is just $\mu_{i}|T_{i}$ plus
standard normal noise, this result implies that the parameters governing
$\mu_{i}|T_{i}$ are strongly identified, leading to the relatively
narrow confidence bounds in Table \ref{tab:estimates}.

Returning to Table \ref{tab:estimates}, the probability of publishing
a marginally-significant t-stat ($\eta$) tends to be larger than
50\% in its bootstrapped distribution, implying that the asset pricing
literature does not strongly discriminate against predictors with
t-stats between 1.96 and 2.58. This result is intuitive, as the asset
pricing literature contains relatively little discussion of marginal
significance, and instead focuses on economic mechanisms.

Table \ref{tab:estimates} shows that the bootstrapped $\eta$ often
bumps against the upper bound of $2/3$ imposed by the model (Equation
(\ref{eq:emp-haircut-support})). This restriction was imposed to
exclude the possibility that weak identification is due to an excessively
general model. Section \ref{sec:robust-pubprob} shows that relaxing
these bounds does not change the main results.

As very similar estimates obtain using either the non-parametric or
semi-parametric bootstrap, for the remainder of the paper I discuss
only the non-parametric bootstrap.

\subsection{Weak Identification of t-stat Hurdles for Cross-Sectional Predictors\label{sec:emp-hurdle}}

With bootstrapped parameters in hand, I can finally quantify weak
and strong identification of multiple testing statistics.

Figure \ref{fig:boot-hurdles} plots the bootstrapped distribution
of t-hurdles (Equation (\ref{eq:theory-hurdle})). Panel (a) examines
the more common FDR upper bound of 5\%, which results in a highly
dispersed t-stat hurdle of between 0 and 3.0. In other words, the
data say little about how to adjust t-hurdles for multiple testing.
Indeed, the classical 5\% counterpart of 1.96 lies right in the middle
of the distribution, implying substantial uncertainty about whether
the hurdle should be raised or lowered.
\noindent \begin{center}
{[}Figure \ref{fig:boot-hurdles} about here{]}
\par\end{center}

The estimates imply a large mode in t-hurdles close to 0, reflective
of the large mode in bootstrapped estimates of $\pi_{F}$ close to
zero that is implicit in Table \ref{tab:estimates}. This mode is
intuitive given the shape of the empirical distribution (Figure \ref{fig:fit}):
This unimodal shape provides little evidence in support of a bimodal
distribution in $\mu_{i}$, and thus simply assuming $\mu_{i}$ is
lognormal ($\pi_{F}=0$) provides a strong fit to the data. This result
is also consistent with Proposition \ref{prop:strong} and Figure
\ref{fig:hlz}, which show that the observed distribution of the full
model is approximately the same as a model with $\pi_{F}=0$.

Some readers may find $\pi_{F}=0$ to be implausible, but Figure \ref{fig:boot-hurdles}
suggests that this mode is immaterial. Excluding the mode at 0, the
distribution of t-hurdle estimates is still highly dispersed, with
values ranging from 1.0 to 3.0. Additionally, Section \ref{sec:robust-pif-gt}
shows that restricting $\pi_{F}\ge0.20$ still results in substantial
uncertainty about the right t-hurdle.

Panel (b) shows that weak identification is also seen if one selects
the unusually restrictive FDR $\le$ 1\% t-hurdle. The resulting distribution
of t-hurdles is, once again, highly dispersed, ranging from 0 to 3.5.
The classical 1\% hurdle of 2.58 lies well-within the interior of
this distribution.

\subsection{Strong Identification of Shrinkage and the Local FDR \label{sec:emp-shrink}}

Figure \ref{fig:boot-bias.pub-fdr} shows that the data are informative
about shrinkage and the local FDR for published predictors. Panel
(a) examines shrinkage by evaluating Equation (\ref{eq:theory-shrink-def})
using bootstrapped data and then taking the mean across published
t-stats within each bootstrap. The panel shows the distribution of
this mean across bootstraps. The vast majority of the distribution
lies below 26\%, which is the upper bound on statistical bias estimated
in \citet{mclean2016does}'s out-of-sample tests. This result implies
that published sample mean returns are at least 74\% due to true expected
returns rather than publication bias, multiple testing, or other related
statistical effects.
\noindent \begin{center}
{[}Figure \ref{fig:boot-bias.pub-fdr} about here{]}
\par\end{center}

The mode of the shrinkage distribution is close to 12\%, which is
the point estimate found by \citet{chen2020publication}. This similarity
obtains despite the fact that I assume a bi-modal distribution in
t-stats, providing a counterexample to \citet{harvey2021uncovering}'s
claim that Chen and Zimmermann's estimate is sensitive to their unimodal
assumption. Indeed, Section \ref{sec:emp-robust} shows Chen and Zimmermann's
estimate is robust to 10 alternative models, all of which use bi-modal
distributions.

Panel (b) shows similar results for the mean local FDR. Panel (b)
computes the mean local FDR by evaluating Equation (\ref{eq:theory-localfdr})
at bootstrapped parameter values and taking the mean across published
t-stats within the bootstrap. Quantitatively similar to Panel (a),
Panel (b) finds that published factors are at least 75\% true, with
high confidence. The figure displays a large mode near 0, due to the
unimodal shape of the empirical data (see Section \ref{sec:emp-hurdle}),
but even if this mode were excluded the estimates still imply that
the published predictors are largely true with high confidence.

For comparison, Figure \ref{fig:boot-bias.pub-fdr} also plots the
mean local FDR for published predictors implied by HLZ's baseline
estimates. HLZ's estimate implies that only 6\% of published findings
are false, close to the middle of the bootstrapped distribution. This
result is surprising, given HLZ's verbal statement that ``most claimed
research findings in financial economics are likely false.'' However,
this estimate is natural given their numerical estimates.

To understand HLZ's numerical estimates, note that the mean local
FDR for published factors is approximately the Bayesian expression
\begin{align}
\E\left[\text{fdr}\left(t_{i}\right)|\pub_{i}\right] & \approx\Pr\left(F_{i}\left||t_{i}|>1.96\right.\right)\nonumber \\
 & =\frac{\Pr\left(|t_{i}|>1.96|F_{i}\right)}{\Pr\left(|t_{i}|>1.96\right)}\pi_{F}.\label{eq:emp-Efdr-approx}
\end{align}
HLZ find that $\pi_{F}=0.444$ (Table 2) and that of 1,378 total t-stats
(also Table 2), 353 t-stats that exceed 1.96 (page 28). Plugging these
numbers into Equation (\ref{eq:emp-Efdr-approx}) leads to 
\begin{align*}
\E\left[\text{fdr}\left(t_{i}\right)|\pub_{i}\right] & \approx\frac{5\%}{353/1378}0.444\\
 & =8.6\%.
\end{align*}
This approximation is a touch higher than the 6\% found by applying
Equation (\ref{eq:theory-localfdr}) to their model simulation because
the simulation assumes a more stringent publication criteria (only
half of t-stats between 1.96 and 2.58 are published). \citet{chen2021most}
finds that even HLZ's conservative estimates imply the FDR is rather
small.

\section{Robustness\label{sec:emp-robust}}

Figures \ref{fig:boot-hurdles} and \ref{fig:boot-bias.pub-fdr} depend
on functional form and distributional assumptions. However, Propositions
\ref{prop:weak} and \ref{prop:strong} do not, suggesting that similar
results may obtain under a wide variety of assumptions.

This subsection confirms this robustness for 10 alternative sets of
assumptions. Table \ref{tab:robust} shows the 5th and 95th percentile
of multiple testing statistics, computed across 500 bootstrapped estimates
for each set of assumptions. Most of the 90\% confidence intervals
for t-hurdles span 0 and 2.9, implying substantial uncertainty about
the proper correction to the classical hurdle of 1.96. In contrast,
all specifications imply shrinkage is at most 29\%, and the local
FDR is at most 25\%, with 95\% confidence.
\noindent \begin{center}
{[}Table \ref{tab:robust} about here{]}
\par\end{center}

Indeed, previous versions of this paper found similar results for
additional estimation and modeling assumptions, including estimations
that account for standard errors and t-stats separately, and estimations
that modeled the full distribution of correlations. Additional robustness
can be found using the github code ({[}a github site{]}) or using
the github code for the previous draft ({[}a github site{]})

The remainder of this section briefly describes each set of assumptions
and their motivations.

\subsection{Robustness to Using Small Chen-Zimmerman t-stats\label{sec:robust-cz-raw}}

Section \ref{sec:emp_data} simplifies modeling by dropping the 24
t-stats that fall below 1.96. Specifications (2) and (3) include these
small t-stats\textemdash as long as they exceed 0.50. The three t-stats
that fall below 0.50 are 0.06, 0.09, and 0.40. Including these tiny
t-stats leads to shrinkage estimates that divide by tiny numbers (Equation
(\ref{eq:theory-shrink-def})), leading to extreme outliers that drive
the means.

To accommodate the additional 21 t-stats that fall below 1.96, I add
two additional steps to the staircase publication probability: 
\begin{align}
\Pr\left(\pub_{i}\right) & =\begin{cases}
\eta_{a}\bar{s} & \hspace{1ex}\text{if }|t_{i}|\le1.50\\
\eta_{b}\bar{s} & \:\text{if }|t_{i}|\in(1.50,1.96]\\
\eta_{c}\bar{s} & \hspace{1ex}\text{if }|t_{i}|\in(1.96,2.58]\\
\bar{s} & \hspace{1ex}\text{if }|t_{i}|>2.58.
\end{cases}\label{eq:emp-pub-prob-1}
\end{align}
where the additional cutoff of 1.50 is chosen to match the cutoff
used in \citet{mclean2016does}. 1.50 is also close to the two-sided
10\% hurdle of 1.64.

Specifications (2) and (3) differ in how they restrict the parameters
$\eta_{a}$, $\eta_{b}$, and $\eta_{c}$. Specification (2) makes
no restrictions. Specification (3) assumes $\eta_{c}\in[1/3,2/3]$
(same as in the baseline), and imposes that the other probabilities
are smaller than 1/3.

\subsection{Robustness to Assuming $\pi_{F}$ is at Least $X\%$\label{sec:robust-pif-gt}}

The baseline estimates show a large mode at $\hat{\pi}_{F}=0.01$
(Table \ref{tab:estimates}), leading to large modes in Figures \ref{fig:boot-hurdles}
and \ref{fig:boot-bias.pub-fdr}. Some readers may find such a small
$\pi_{F}$ implausible and would like to restrict $\pi_{F}$ to some
minimum value.

In Table \ref{tab:robust}, Specifications (4) and (5) use the restrictions
$\pi_{F}\ge0.10$ and $\pi_{F}\ge0.20$, respectively.

\subsection{Robustness to Alternative Publication Probability Functions\label{sec:robust-pubprob}}

The baseline model restricts $\eta\in[1/3,2/3]$ to show that weak
identification is not due to an excessively general model. Specification
(6) restricts even further to $\eta=0.5$, which is HLZ's baseline
assumption.

Specification (7) examines a more general $\eta\in[1/3,1.0]$. This
choice is motivated by the fact that the baseline bootstrap bumps
up against the upper bound of $\eta=$2/3 (Table \ref{tab:estimates}
), suggesting that $\eta\ge2/3$ is preferred by the data.

Specification (8) assumes a logistic form for the publication probability
\begin{align*}
\Pr\left(\pub_{i}\right) & =\frac{1}{1+\exp\left(-\eta_{b}(|t_{i}|-\eta_{a})\right)},
\end{align*}
which is the same function used in \citet{chen2020publication} (see
also \citet{cochrane2005risk,harvey2021uncovering}).

\subsection{Robustness to Alternative Distributional Assumptions\label{sec:robust-mudist}}

The baseline model assumes $\mu_{i}|T_{i}$ is log-normal, as this
is the simplest assumption one can make that ensures that true predictors
have a positive mean return that is distinct from 0.

In Table \ref{tab:robust}, Specification (9) assumes $\mu_{i}|T_{i}$
is exponential following HLZ and Specification (10) assumes $\mu_{i}|T_{i}$
is a scaled t-distribution, similar to \citet{andrews2019identification}
and \citet{chen2020publication}. Specification (11) assumes $\mu_{i}|T_{i}$
is a mixture of two normals, which can be motivated by the idea that
there are two distinct types of true predictors, each with a bell-shaped
distribution.

Specifications (10) and (11) often imply that $\mu_{i}|T_{i}$ can
be either positive or negative, which implies that the sign of $t_{i}$
should be accounted for in the shrinkage estimation (Equation (\ref{eq:theory-shrink-def})).
For these two rows, the shrinkage shown thus uses
\begin{align}
\text{shrinkage}_{\text{signed}}\left(\bar{t}\right) & \equiv\frac{\bar{t}-E\left(\mu_{i}\left|t_{i}=\bar{t}\right.\right)}{\bar{t}}.\label{eq:theory-bias-1}
\end{align}
which implicitly assumes that the journals use theory to find the
appropriate sign of predictability. Most papers in the Chen-Zimmermann
dataset motivate their signs using theory.

\section{Conclusion}

Motivated by concerns about scientific credibility, a series of papers
call for statistical hurdles to be raised. I show these calls may
be difficult to justify empirically. Publication bias means that the
key parameter in these arguments may be weakly identified. As a result,
empirical data may say little about whether hurdles should be raised,
stay the same, or even be lowered.

More positively, I find that other multiple testing statistics can
be strongly identified. In particular, shrinkage and the local FDR
for published factors may be able to be pinned down without knowledge
of t-stats near zero. For the cross-sectional return predictability
literature, these statistics imply that published findings are at
least 75\% true, with high confidence, across many model specifications.

What ``at least 75\% true'' says about the proper $t$-hurdle is
subjective, but at least some readers will argue that a literature
that is mostly true is a healthy one. Still, others may argue that
a literature should strive for trust, and even for achieving ``no
questions asked'' from their readers.

A caveat about these empirical results is that they are based on the
field of cross-sectional predictability. While other fields of economics
also feature very large t-stats (\citet{brodeur2016star}), a rigorous
analysis is required to pin down the FDR. Moreover, cross-sectional
predictability is known for its clean data, standardized methods,
and strong replicability (\citet{ChenZimmermann2021,jensen2021there}).
These features allay concerns about coding errors and the improper
calculation of t-stats that may remain if shrinkage and FDR estimates
were applied to other fields.

Regardless, my results demonstrate that the debate about scientific
credibility should take multiple testing statistics more seriously.
These statistics do \emph{not} simply say that statistical hurdles
should be raised, or that a large share of findings are false. Identification
is important, and some multiple testing statistics are more strongly
identified than others. Most important, my paper shows that the debate
should focus on the more strongly identified statistics.

\newpage{}

\appendix

\section{A Model with Supporting Evidence\label{sec:app-zevidence}}

\setcounter{table}{0} \renewcommand{\thetable}{A.\arabic{table}}
\setcounter{figure}{0} \renewcommand{\thefigure}{A.\arabic{figure}}

This section presents a simple model for analyzing the bias the comes
from ignoring supporting evidence.

As in Section \ref{sec:theory-model}, the t-stat $t_{i}$ provides
a noisy signal of the latent effect $\mu_{i}$, but for simplicity
assume $\mu_{i}$ is normal: 
\begin{align*}
t_{i}|\mu_{i} & \sim\text{Normal}\left(\mu_{i},1\right)\\
\mu_{i} & \sim\text{Normal}\left(0,\lambda\right)
\end{align*}
 where $\lambda$ in this setting is the variance of $\mu_{i}$. This
is the key parameter that determines shrinkage.

Add to the Section \ref{sec:theory-model} model supporting evidence,
represented by $z_{i}$:
\begin{align*}
z_{i}|\mu_{i} & \sim\text{Normal}\left(\mu_{i,}1\right)\\
\text{\Cov}\left(t_{i},z_{i}|\mu_{i}\right) & =\rho
\end{align*}
where $\rho$ allows for correlation between the noise components
of $t_{i}$ and $z_{i}$. One can think of $z_{i}$ as representing
aspects of a research paper that are not captured by the primary t-stat.

Now, extend Equation (\ref{eq:theory-pubprob}) to allow publication
to depend on $z_{i}$: 
\begin{align*}
\Pr\left(\pub_{i}\left|t_{i},z_{i}\right.\right) & =\begin{cases}
\bar{s} & \left(t_{i}>2.5\right)\cap\left(z_{i}>z_{\min}\right)\\
0 & \text{otherwise}
\end{cases}
\end{align*}
where the strict cutoffs and the lack of absolute values on $t_{i}$
are chosen for simplicity. The key parameter in this expression is
$z_{\min}$, which captures the quality of supporting evidence required
by the literature.

Let $\hat{\lambda}$ be the following estimate that ignores the supporting
evidence:
\begin{align}
\widehat{\lambda} & =\arg\min_{\lambda}\left(N_{\pub}^{-1}\sum_{i:\pub_{i}}t_{i}-E\left(t_{i}|t_{i}>2.5;\lambda\right)\right).\label{eq:app-lambdahat}
\end{align}
This expression formalizes the misspecification error. The correct
model moment is $E\left(t_{i}|t_{i}>2.5,z_{i}>z_{\min};\lambda\right)$,
not $E\left(t_{i}|t_{i}>2.5;\lambda\right)$.

\begin{figure}[h!]
\caption{\textbf{Supporting Evidence and Bias in $\hat{\lambda}$}}
\label{fig:app-zevidence-lambda} Model is described in Appendix \ref{sec:app-zevidence},
with $\rho=0$. Grey is $E\left(t_{i}|t_{i}>2.5;\lambda\right)$ as
a function of $\lambda$ and red is $E\left(t_{i}|t_{i}>2.5,z_{i}>1;\lambda\right)$.
The ``Data'' line and the choice of $z_{\min}=1.0$ as ``actual''
are arbitrary and selected for illustration. \vspace{0.15in}

\centering
\includegraphics[width=0.9\textwidth]{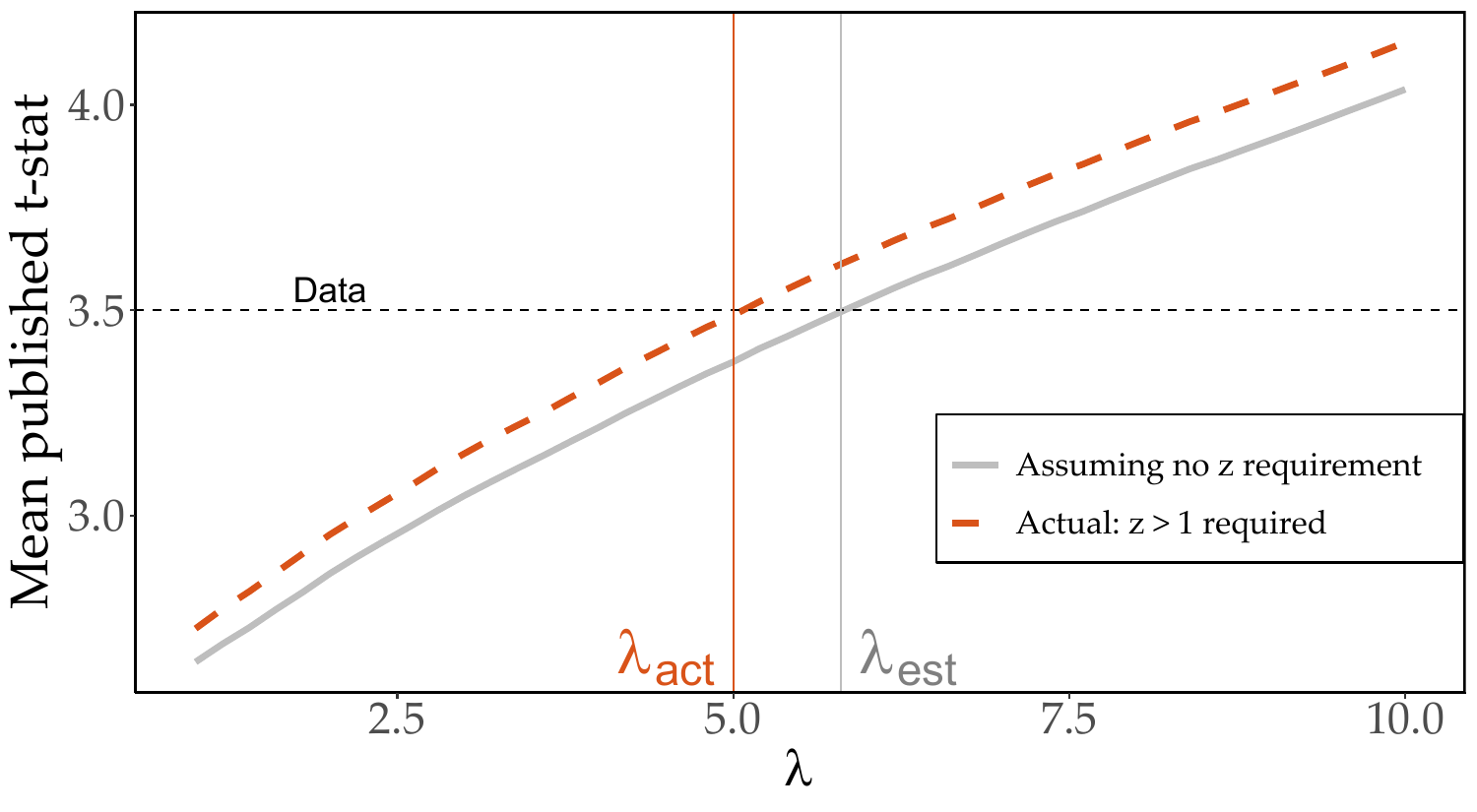} 
\end{figure}

In general, $\widehat{\lambda}$ is upward biased. This happens because
the target moment $E\left(t_{i}|t_{i}>2.5,z_{i}>z_{\min};\lambda\right)$
is increasing in $z_{\min}$, due to the fact that both $t_{i}$ and
$z_{i}$ are positively related to $\mu_{i}$. Figure \ref{fig:app-zevidence-lambda}
visualizes this bias. An upward-biased $\hat{\lambda}$ implies a
downward biased shrinkage.

There is, however, is an offsetting bias. Ignoring supporting evidence
in the empirical Bayes estimate of $\mu_{i}$ leads to
\begin{align}
E\left(\mu_{i}|t_{i};\lambda\right) & <E\left(\mu_{i}|t_{i},z_{i};\lambda\right)\quad\text{for }z_{i}>z_{\min}\label{eq:app-muhat}
\end{align}
since $z_{i}$ is positively related to $\mu_{i}$ and published data
tend to have high $z_{i}$. A downward bias in estimates of $\mu_{i}$
leads to an upward bias in shrinkage.

Which bias dominates is not immediately clear. In a simulation analysis,
I find the net bias can go either way, though the magnitudes of the
bias are generally moderate. I examine many simulations with values
of $\lambda$ ranging from 0.5 to 12, $\rho$ from 0 to 0.6, and $z_{\min}$
from 0 to 2.

\begin{figure}[h!]
\caption{\textbf{Net Bias in Shrinkage under Various Parameter Values}}
\label{fig:app-zevidence-shrink} I simulate the model in Appendix
\ref{sec:app-zevidence} under various parameter values, estimate
$\hat{\lambda}$ ignoring the supporting evidence, and then estimate
shrinkage. \vspace{0.15in}

\centering
\includegraphics[width=0.95\textwidth]{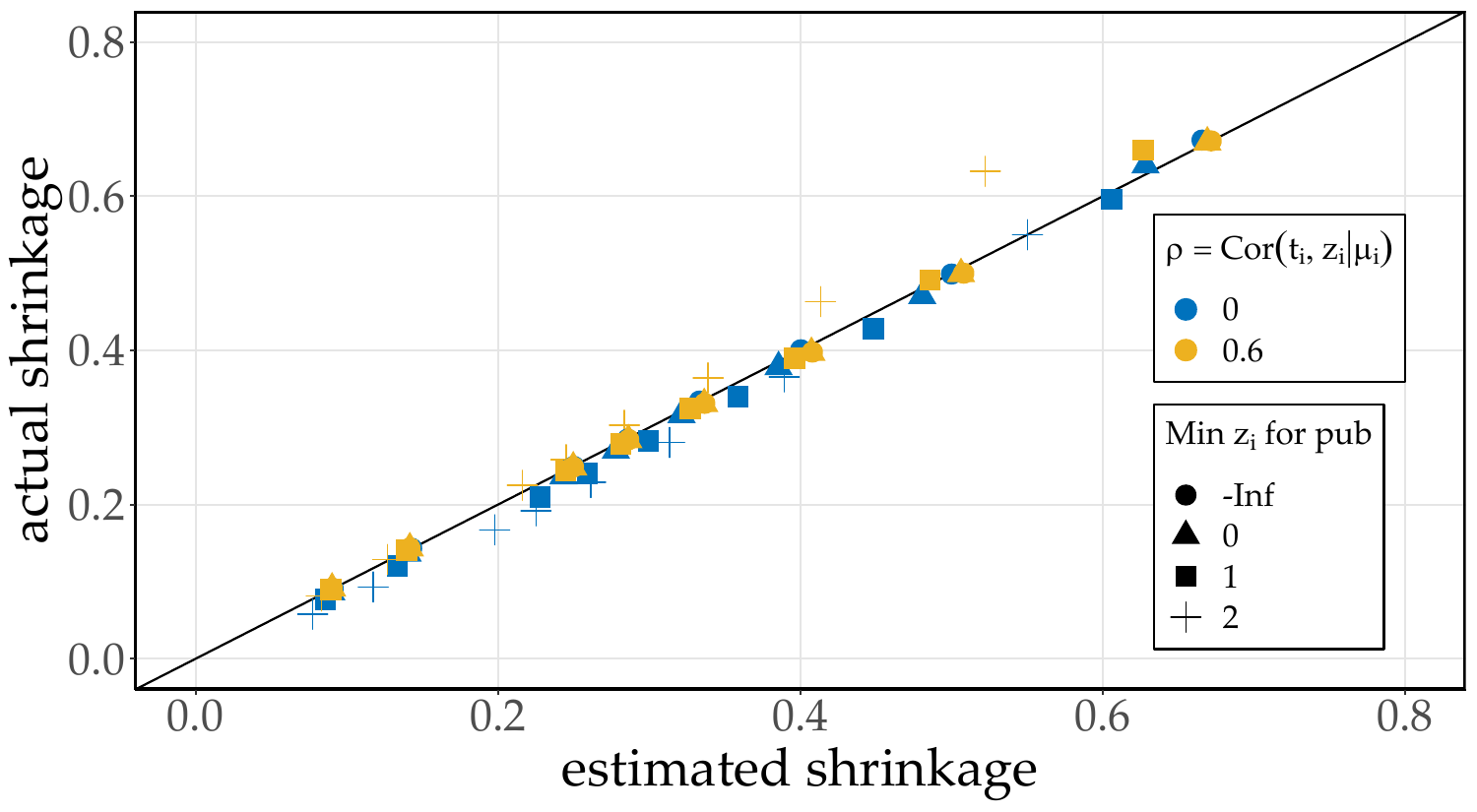} 
\end{figure}

Figure \ref{fig:app-zevidence-shrink} presents the simulation results.
In each simulation, I estimate $\hat{\lambda}$ using the potentially
misspecified Equation (\ref{eq:app-lambdahat}), compute the potentially
mis-specified empirical Bayes estimate of $\mu_{i}$ (Equation (\ref{eq:app-muhat})),
and then compute shrinkage by taking the mean estimated $\mu_{i}$
for published $i$ and dividing by the mean $t_{i}$ for published
$i$. Actual shrinkage uses the actual $\mu_{i}$. The actual shrinkage
can be both above and below the estimated shrinkage. But generally
the errors are small relative to the estimated shrinkage.

The errors can become noticeable for $z_{\min}=2.0$, suggesting that
even higher levels of $z_{\min}$ can result in substantial bias.
One can argue, however, that the standards for supporting evidence
should not be as strict as the standards for the main evidence, and
thus $z_{\min}<2.0$ is reasonable.

\section{Proofs}

\subsection{Proof of Proposition \ref{prop:main}\label{sec:app-proof-main}}
\begin{proof}
In this proof, I consider a more general version of the proposition,
in which the FDR-based and traditional hurdles use different critical
levels. For this more general proof, we need to define a general classical
hurdle: 
\begin{align}
h_{0}\left(\alpha_{0}\right) & \equiv\min_{h\in\mathbb{R}_{+}}\left\{ h:\Pr\left(|t_{i}|>h\bigl|F_{i}\right)\le\alpha_{0}\right\} ,\label{eq:theory-single-test}
\end{align}
where $\alpha_{0}$ is the classical significance level. Similarly,
generalize Equation (\ref{eq:theory-hurdle}) as 
\begin{align*}
h_{\FDR}^{\ast}\left(\alpha\right) & \equiv\min_{h\in\mathbb{R}_{+}}\left\{ h:\Pr\left(F_{i}||t_{i}|>h\right)\le\alpha\right\} .
\end{align*}

Since $\Pr\left(F_{i}||t_{i}|>h\right)$ is strictly decreasing, the
solution to the minimization (\ref{eq:theory-hurdle}) solves $\Pr\left(F_{i}||t_{i}|>h_{\FDR}^{\ast}\left(\alpha\right)\right)=\alpha$.
Rewrite this using Bayes rule: 
\begin{align}
\alpha & =\Pr\left(F_{i}||t_{i}|>h_{\FDR}^{\ast}\left(\alpha\right)\right)=\frac{\Pr\left(|t_{i}|>h_{\FDR}^{\ast}\left(\alpha\right)\bigl|F_{i}\right)}{\Pr\left(|t_{i}|>h_{\FDR}^{\ast}\left(\alpha\right)\right)}\pi_{F}\label{eq:app-proof1-01}
\end{align}
Then Equation (\ref{eq:theory-single-test}) is solved with Pr $\left(|t_{i}|>h_{0}\left(\alpha_{0}\right)|F_{i}\right)=\alpha_{0}$.
Multiplying both sides by Equation (\ref{eq:app-proof1-01}) we have
\begin{align}
\Pr\left(|t_{i}|>h_{0}\left(\alpha_{0}\right)|F_{i}\right) & =\Pr\left(|t_{i}|>h_{\FDR}^{\ast}\left(\alpha\right)\bigl|F_{i}\right)\left[\frac{\alpha_{0}}{\alpha}\frac{\pi_{F}}{\Pr\left(|t_{i}|>h_{\FDR}^{\ast}\left(\alpha\right)\right)}\right]\label{eq:app-main-1}
\end{align}
From here, we have three cases for the square bracket term, but the
first case is symmetric with the rest.

For the first case, suppose 
\begin{align*}
1 & >\left[\frac{\alpha}{\alpha_{0}}\frac{\Pr\left(|t_{i}|>h_{\FDR}^{\ast}\left(\alpha\right)\right)}{\pi_{F}}\right]
\end{align*}
Note that the above expression is equivalent to the condition in Equation
(\ref{eq:theory-prop-result-1}), if we use $\alpha=\alpha_{0}=5\%$.
Multiply both sides of this inequality with (\ref{eq:app-main-1})
to get the equivalent expression 
\begin{align*}
Pr\left(|t_{i}|>h_{0}\left(\alpha_{0}\right)|F_{i}\right) & >\Pr\left(|t_{i}|>h_{\FDR}^{\ast}\left(\alpha\right)\bigl|F_{i}\right).
\end{align*}
Since $Pr\left(|t_{i}|>h|F_{i}\right)$ strictly decreases in $h$,
the above equation is equivalent to 
\begin{align*}
h_{0}\left(\alpha_{0}\right) & <h_{\FDR}^{\ast}\left(\alpha\right).
\end{align*}
Thus we have shown the equivalence in Equation (\ref{eq:theory-prop-result-1}).

For the other two cases, simply replace $>$ with either $=$ or $<$
in the previous paragraph.
\end{proof}

\subsection{Proof of Proposition \ref{prop:weak}\label{sec:app-proof-weak}}
\begin{proof}
For ease of notation, define 
\begin{align*}
\delta\left(X\right) & \equiv\frac{\Pr\left(X\bigl|F_{i}\right)}{\Pr\left(X\bigl|T_{i}\right)}\\
L\left(\pi\right) & \equiv\frac{\pi}{(1-\pi)}
\end{align*}
where $X$ is some event. Note that for $X=|t_{i}|\subset[\tgood,\infty)$
we have by assumption (\ref{eq:theory-weak-cond})
\begin{align*}
\delta\left(X\right) & <\varepsilon.
\end{align*}

Use this notation to rewrite Equation (\ref{eq:theory-cond-CDF}):
\begin{align}
\Pr\left(|t_{i}|\le\bar{t}||t_{i}|>\tgood\right) & =\frac{\Pr\left(|t_{i}|\in(\tgood,\bar{t}]\bigl|T_{i}\right)}{\Pr\left(|t_{i}|>\tgood\bigl|T_{i}\right)}\left[\frac{1+L\left(\pi_{F}\right)\delta\left(|t_{i}|\in(\tgood,\bar{t}]\right)}{1+L\left(\pi_{F}\right)\delta\left(|t_{i}|>\tgood\right)}\right]\label{eq:app-key}
\end{align}
where all I did was factor $(1-\pi_{F})$ and the probabilities that
condition on $T_{i}$. In the above equation, the front term doesn't
depend on $\pi_{F}$. Thus, if we take the ratio of this expression
evaluated at two different values of $\pi_{F}$, the first term cancels
out:
\begin{align}
\frac{\Pr\left(|t_{i}|\le\bar{t}||t_{i}|>\tgood\right)|_{\pi_{F}=\pi'}}{\Pr\left(|t_{i}|\le\bar{t}||t_{i}|>\tgood\right)|_{\pi_{F}=\pi}} & =\left[\frac{1+L\left(\pi'\right)\delta\left(|t_{i}|\in(\tgood,\bar{t}]\right)}{1+L\left(\pi'\right)\delta\left(|t_{i}|>\tgood\right)}\right]\left[\frac{1+L\left(\pi\right)\delta\left(|t_{i}|>\tgood\right)}{1+L\left(\pi\right)\delta\left(|t_{i}|\in(\tgood,\bar{t}]\right)}\right]\nonumber \\
 & =\left[\frac{1+L\left(\pi'\right)\delta\left(|t_{i}|\in(\tgood,\bar{t}]\right)}{1+L\left(\pi\right)\delta\left(|t_{i}|\in(\tgood,\bar{t}]\right)}\right]\left[\frac{1+L\left(\pi\right)\delta\left(|t_{i}|>\tgood\right)}{1+L\left(\pi'\right)\delta\left(|t_{i}|>\tgood\right)}\right]\label{eq:app-prop2-01}\\
 & =1+\left[L\left(\pi'\right)-L\left(\pi\right)\right]\left[\delta\left(|t_{i}|\in(\tgood,\bar{t}]\right)-\delta\left(|t_{i}|>\tgood\right)\right]+O\left(\varepsilon^{2}\right)\label{eq:app-prop2-02}
\end{align}
where the last line uses a Taylor expansion around $\left[\delta\left(|t_{i}|\in(\tgood,\bar{t}]\right),\delta\left(|t_{i}|>\tgood\right)\right]=\left[0,0\right]$
and the fact $\delta\left((\tgood,\bar{t}]\right)<\varepsilon$ and
$\delta\left(|t_{i}|>\tgood\right)<\varepsilon$. The simplicity of
this Taylor expansion comes from the fact that 
\begin{align*}
\left.\frac{d}{dx}\left(\frac{1+ax}{1+bx}\right)\right|_{x=0} & =\left.\frac{a}{bx+1}-\frac{b(ax+1)}{(bx+1)^{2}}\right|_{x=0}=a-b.
\end{align*}
So the partial derivatives of Equation (\ref{eq:app-prop2-01}) simplify
dramatically once $\left[\delta\left(|t_{i}|\in(\tgood,\bar{t}]\right),\delta\left(|t_{i}|>\tgood\right)\right]=\left[0,0\right]$
is plugged in.

Subtracting 1 form both sides, taking absolute values of (\ref{eq:app-prop2-02})
and applying the triangle inequality
\begin{align*}
\left|\frac{\Pr\left(|t_{i}|\le\bar{t}||t_{i}|>\tgood\right)|_{\pi_{F}=\pi'}}{\Pr\left(|t_{i}|\le\bar{t}||t_{i}|>\tgood\right)|_{\pi_{F}=\pi}}-1\right| & \le\left|L\left(\pi'\right)-L\left(\pi\right)\right|\left|\delta\left(|t_{i}|\in(\tgood,\bar{t}]\right)-\delta\left(|t_{i}|>\tgood\right)\right|+O\left(\varepsilon^{2}\right)\\
 & \le\left|L\left(\pi'\right)-L\left(\pi\right)\right|\varepsilon+O\left(\varepsilon^{2}\right)\\
 & \le\frac{\bar{\pi}}{1-\bar{\pi}}\varepsilon+O\left(\varepsilon^{2}\right).
\end{align*}
Recall $\bar{\pi}$ is the maximum value that $\pi$ and $\pi'$ can
take. The third line uses the fact that $L\left(\pi\right)$ is strictly
increasing.

Finally, imposing the assumption that $\bar{\pi}=2/3$ finishes the
proof. Note that using instead $\bar{\pi}=1/2$ results in a factor
of 1 in front of $\varepsilon$, while $\bar{\pi}=9/10$ would result
in a factor of 9. $\bar{\pi}=2/3$ was chosen as an illustrative middle
ground.
\end{proof}

\subsection{Derivation of Shrinkage Expression (\ref{eq:theory-bias-approx})
\label{sec:app-bias-approx}}

Since $i$ is used as a subscript throughout, I drop it for ease of
notation. I use the shorthand notation $p\left(\mu|t\right)$ to denote
the density of $\mu$ given $|t|$, $p\left(\mu|T\right)$ to denote
the density of $\mu$ given the factor is true, and similarly for
other random variables.

Write down the posterior of $\mu|t$ and rewrite the denominator using
the law of total probability
\begin{align*}
p\left(\mu|t\right) & =\frac{p\left(t|\mu\right)p\left(\mu|T\right)\pi_{T}+p\left(t|\mu\right)p\left(\mu|F\right)\pi_{F}}{p\left(t\right)}\\
 & =\frac{p\left(t|\mu\right)p\left(\mu|T\right)\pi_{T}+p\left(t|\mu\right)p\left(\mu|F\right)\pi_{F}}{p\left(t|T\right)\pi_{T}+p\left(t|F\right)\pi_{F}}
\end{align*}
Then factor out $p\left(t|T\right)\pi_{T}$ 
\begin{align*}
p\left(\mu|t\right) & =\left[p\left(t|\mu\right)\frac{p\left(\mu|T\right)}{p\left(t|T\right)}+p\left(t|\mu\right)\frac{p\left(\mu|F\right)\pi_{F}}{p\left(t|T\right)\pi_{T}}\right]\\
 & \quad\times\left\{ \frac{1}{1+\frac{p\left(t|F\right)\pi_{F}}{p\left(t|T\right)\pi_{T}}.}\right\} \\
 & =\left[p\left(\mu|t,T\right)+p\left(t|\mu\right)\frac{p\left(\mu|F\right)\pi_{F}}{p\left(t|T\right)\pi_{T}}\right]\left\{ \frac{1}{1+\Delta\left(t,\pi_{F},\lambda\right)}\right\} 
\end{align*}
where 
\begin{align*}
\Delta\left(t,\pi_{F},\lambda\right) & \equiv\frac{p\left(t|F\right)\pi_{F}}{p\left(t|T\right)\pi_{T}},
\end{align*}
since $p\left(t|\mu,T\right)=p\left(t|\mu\right)$.

Let $h\left(\mu\right)$ be a function s.t. $h\left(0\right)=0$.
Then the posterior expected value of $h\left(\mu\right)$ is 
\begin{align}
E\left(h\left(\mu\right)|t,\pi_{F},\lambda\right) & =\left\{ E\left(h\left(\mu\right)|t,T,\lambda\right)+\int d\mu h\left(\mu\right)p\left(t|\mu\right)\frac{p\left(\mu|F\right)\pi_{F}}{p\left(t|T\right)\pi_{T}}\right\} \left\{ \frac{1}{1+\Delta\left(t,\pi_{F},\lambda\right)}\right\} \nonumber \\
 & =\left\{ E\left(h\left(\mu\right)|t,T,\lambda\right)+h\left(0\right)p\left(t|0\right)\frac{\pi_{F}}{p\left(t|T\right)\pi_{T}}\right\} \left\{ \frac{1}{1+\Delta\left(t,\pi_{F},\lambda\right)}\right\} \nonumber \\
 & =E\left(h\left(\mu\right)|t,T,\lambda\right)\left\{ \frac{1}{1+\Delta\left(t,\pi_{F},\lambda\right)}\right\} \nonumber \\
 & =E\left(h\left(\mu\right)|t,\pi_{F}=0,\lambda\right)\left\{ \frac{1}{1+\Delta\left(t,\pi_{F},\lambda\right)}\right\} \label{eq:app-Ehmu}
\end{align}
where the second line comes from the fact that $p\left(\mu|F\right)=\delta\left(\mu\right)$,
where $\delta\left(\cdot\right)$ is the Dirac delta, the third uses
$h\left(0\right)=0$, and the last line uses the fact that the model
with $\pi_{F}=0$ assumes that all factors are true. Choosing $h\left(\mu\right)=\mu$
leads to Equation (\ref{eq:theory-bias-approx}).

A similar argument can be made for the local FDR. The local FDR can
be rewritten as 
\begin{align*}
\fdr\left(t\right) & =\frac{p\left(t|F\right)\pi_{F}}{p\left(t|T\right)\pi_{T}+p\left(t|F\right)\pi_{F}}\\
 & \le\frac{p\left(t|F\right)}{p\left(t|T\right)\pi_{T}+p\left(t|F\right)\pi_{F}}\\
 & =\frac{p\left(t|F\right)}{p\left(t|T\right)}\left[\frac{1}{1+\frac{p\left(t|F\right)\pi_{F}}{p\left(t|T\right)}}\right].
\end{align*}
$\frac{p\left(t|F\right)}{p\left(t|T\right)}$ is, in a way, a bound
on the local fdr assuming that all factors are true. The square brackets
is the correction term, which is just $\Delta\left(t,\pi_{F},\lambda\right)\times\pi_{T}$.

\subsection{Proof of Proposition \ref{prop:strong}\label{sec:app-proof-strong}}
\begin{proof}
The proposition uses the same assumptions as Proposition \ref{prop:weak},
so I can borrow this expression from its proof 
\begin{align*}
\Pr\left(|t_{i}|\le\bar{t}||t_{i}|>\tgood\right) & =\frac{\Pr\left(|t_{i}|\in(\tgood,\bar{t}]\bigl|T_{i}\right)}{\Pr\left(|t_{i}|>\tgood\bigl|T_{i}\right)}\left[\frac{1+L\left(\pi_{F}\right)\delta\left(|t_{i}|\in(\tgood,\bar{t}]\right)}{1+L\left(\pi_{F}\right)\delta\left(|t_{i}|>\tgood\right)}\right]
\end{align*}
where
\begin{align*}
\delta\left(X\right) & \equiv\frac{\Pr\left(X\bigl|F_{i}\right)}{\Pr\left(X\bigl|T_{i}\right)}\\
L\left(\pi\right) & \equiv\frac{\pi}{(1-\pi)}
\end{align*}
and $X$ is some event. Rearranging, 
\begin{align*}
\frac{\Pr\left(|t_{i}|\le\bar{t}||t_{i}|>\tgood\right)}{\Pr\left(|t_{i}|\in(\tgood,\bar{t}]\bigl|T_{i},t_{i}|>\tgood\right)} & =\left[\frac{1+L\left(\pi_{F}\right)\delta\left(|t_{i}|\in(\tgood,\bar{t}]\right)}{1+L\left(\pi_{F}\right)\delta\left(|t_{i}|>\tgood\right)}\right]\\
 & =1+L\left(\pi_{F}\right)\left[\delta\left(|t_{i}|\in(\tgood,\bar{t}]\right)-\delta\left(|t_{i}|>\tgood\right)\right]+O\left(\varepsilon^{2}\right)
\end{align*}
or 
\begin{align*}
\frac{\Pr\left(|t_{i}|\le\bar{t}||t_{i}|>\tgood\right)}{\Pr\left(|t_{i}|\in(\tgood,\bar{t}]\bigl|T_{i},t_{i}|>\tgood\right)}-1 & =L\left(\pi_{F}\right)\left[\delta\left(|t_{i}|\in(\tgood,\bar{t}]\right)-\delta\left(|t_{i}|>\tgood\right)\right]+O\left(\varepsilon^{2}\right)
\end{align*}
Taking absolute values and manipulating
\begin{align*}
\left|\frac{\Pr\left(|t_{i}|\le\bar{t}||t_{i}|>\tgood\right)}{\Pr\left(|t_{i}|\in(\tgood,\bar{t}]\bigl|T_{i},t_{i}|>\tgood\right)}-1\right| & \le\left|L\left(\pi_{F}\right)\right|\left|\left[\delta\left(|t_{i}|\in(\tgood,\bar{t}]\right)-\delta\left(|t_{i}|>\tgood\right)\right]\right|+O\left(\varepsilon^{2}\right)\\
 & \le\left|L\left(\pi_{F}\right)\right|\varepsilon+O\left(\varepsilon^{2}\right)\\
 & \le\frac{\bar{\pi}}{1-\bar{\pi}}\varepsilon+O\left(\varepsilon^{2}\right)
\end{align*}
where $\bar{\pi}$ is the maximum value under consideration for $\pi_{F}$.
The 2nd line comes from the fact that $\delta\left(|t_{i}|>\tgood\right)>0$
and $\delta\left(|t_{i}|\in(\tgood,\bar{t}]\right)<\varepsilon$ and
the third line uses the fact that $L\left(\pi_{F}\right)$ is strictly
increasing. Plugging in $\bar{\pi}=2/3$ completes the proof.
\end{proof}

\subsection{Proof of Corollary \ref{prop:bh-no-answer}\label{sec:app-proof-no-answer}}

Comparing the definition of $\widehat{\FDR}\left(h\right)$ (Equation
(\ref{eq:prop-equiv-FDRhat})) and Equation (\ref{eq:theory-fdr-bayes})
we see that $h_{BY}$ is isomorphic to replacing $\pi_{F}/\Pr\left(|t_{i}|>h\right)$
with $\left(\sum_{i=1}^{N}\frac{1}{i}\right)/\widehat{\Pr}\left(|t_{i}|>h\right)$
in the hurdle definition (Equation (\ref{eq:theory-hurdle})). Thus,
Proposition \ref{prop:main} implies 
\begin{align*}
h_{BY}>1.96\text{ if and only if }\left(\sum_{i=1}^{N}\frac{1}{i}\right)>\widehat{\Pr}\left(|t_{i}|>1.96\right) & .
\end{align*}
But since $\sum_{i=1}^{N}\frac{1}{i}>1$ and $\widehat{\Pr}\left(|t_{i}|>1.96\right)\equiv\frac{\sum_{i=1}^{N}I\left(|t_{i}|>1.96\right)}{N}<1$,
we have $h_{BY}>1.96$.

\section{Estimation Details}

\subsection{QML on Simulated Data with Correlations\label{sec:app-sim-est}}

This section shows that QML works well even if factors have pairwise
correlations as high as 0.9 assuming no publication bias.

The simulation model begin with the model in Section \ref{sec:emp-model}
and specifies correlations to be AR1 across factor indexes:
\begin{align*}
\varepsilon_{i} & \equiv t_{i}-\mu_{i}\\
\varepsilon_{i} & \sim N\left(\rho\varepsilon_{i-1},\sqrt{1-\rho^{2}}\right)
\end{align*}
This specification ensures Equation (\ref{eq:theory-t_mu}) holds
and provides a simple way to model correlated factors.

I assume all factors are observed. One can think of this exercise
as demonstrating that the weak identification I find is due to publication
bias rather than QML.

I fix $\lambda_{\mu}$ and $\lambda_{\sigma}$ using the point estimate
from Table \ref{tab:estimates}. I vary $\pi_{F}$ and $\rho$ as
shown in Table \ref{tab:qml-check}. For each of the shown parameter
values, I simulate a sample of 200 factors, apply QML, and repeat
100 times. The table shows the mean $\hat{\pi}_{F}$ across the 100
simulations as well as the standard deviation across simulations.

The table shows that QML is approximately unbiased for $\pi_{F}$
and has moderate standard errors of around 0.10 for $\rho\le0.5$.
Standard errors increase to about 0.15 for the $\rho=$0.90 cases,
but they are still half as large as the standard error of about 0.33
in the empirical estimates. Overall Table \ref{tab:qml-check} shows
that weak identification of $\pi_{F}$ is not due to QML.

\begin{table}[h!]
\caption{\textbf{QML on Simulated Data with Correlations}}
\label{tab:qml-check} \vspace{0.1in}

{\small{}I simulate factors with correlated t-stats that have AR1
coefficient $\rho$ and apply QML. Results show means and standard
deviations across 100 simulations using each parameter set.}{\small\par}
\begin{centering}
\vspace{0ex}
\par\end{centering}
\centering{}\setlength{\tabcolsep}{4ex} 
\begin{center} 
% ==== paste below

% Table generated by Excel2LaTeX from sheet 'qmlcheck' 

\begin{tabular}{rrrrr} \toprule \multicolumn{1}{c}{\multirow{2}[2]{*}{$\pi_F$ truth}} &   & \multicolumn{3}{c}{Cross-predictor AR1 coefficient $\rho$} \\   &   & \multicolumn{1}{c}{0.1} & \multicolumn{1}{c}{0.5} & \multicolumn{1}{c}{0.9} \\ \cmidrule{1-1}\cmidrule{3-5}\multicolumn{1}{c}{0.10} & \multicolumn{1}{l}{mean $\hat{\pi}_F$} & \multicolumn{1}{c}{          0.13 } & \multicolumn{1}{c}{          0.13 } & \multicolumn{1}{c}{                                                                          0.12 } \\   & \multicolumn{1}{l}{sd $\hat{\pi}_F$} & \multicolumn{1}{c}{          0.10 } & \multicolumn{1}{c}{          0.10 } & \multicolumn{1}{c}{                                                                          0.12 } \\ \multicolumn{1}{c}{0.50} & \multicolumn{1}{l}{mean $\hat{\pi}_F$} & \multicolumn{1}{c}{          0.48 } & \multicolumn{1}{c}{          0.49 } & \multicolumn{1}{c}{                                                                          0.49 } \\   & \multicolumn{1}{l}{sd $\hat{\pi}_F$} & \multicolumn{1}{c}{          0.15 } & \multicolumn{1}{c}{          0.14 } & \multicolumn{1}{c}{                                                                          0.17 } \\ \multicolumn{1}{c}{0.90} & \multicolumn{1}{l}{mean $\hat{\pi}_F$} & \multicolumn{1}{c}{          0.90 } & \multicolumn{1}{c}{          0.89 } & \multicolumn{1}{c}{                                                                          0.87 } \\   & \multicolumn{1}{l}{sd $\hat{\pi}_F$} & \multicolumn{1}{c}{          0.07 } & \multicolumn{1}{c}{          0.08 } & \multicolumn{1}{c}{                                                                          0.15 } \\ \midrule   &   &   &   &  \\ \end{tabular}% 

% ==== end paste
\end{center} 
\end{table}

\subsection{Details on the Semi-Parametric Bootstrap \label{sec:app-semiparboot}}

The cluster-bootstrap uses post-1963 panel returns from the Chen-Zimmermann
data. Unlike the main results which use only in-sample data, I include
data outside of the original sample periods for the cluster bootstrap.
This sample selection helps keep the panel balanced and is conservative
because correlations generally increase post-publication (\citet{mclean2016does}).
Using this sample, I construct residuals by subtracting the mean return
at the predictor level.

\begin{figure}[h!]
\caption{\textbf{Correlations in Cluster-Bootstrapped Return Residuals.}}
\label{fig:semipar-cor} I bootstrap from Chen-Zimmermann panel returns
while ensuring that returns in the same month are always drawn together.
The plot compares the distribution of correlations in the bootstrap
against the correlations in the raw data. \vspace{0.15in}

\centering
\includegraphics[width=0.6\textwidth]{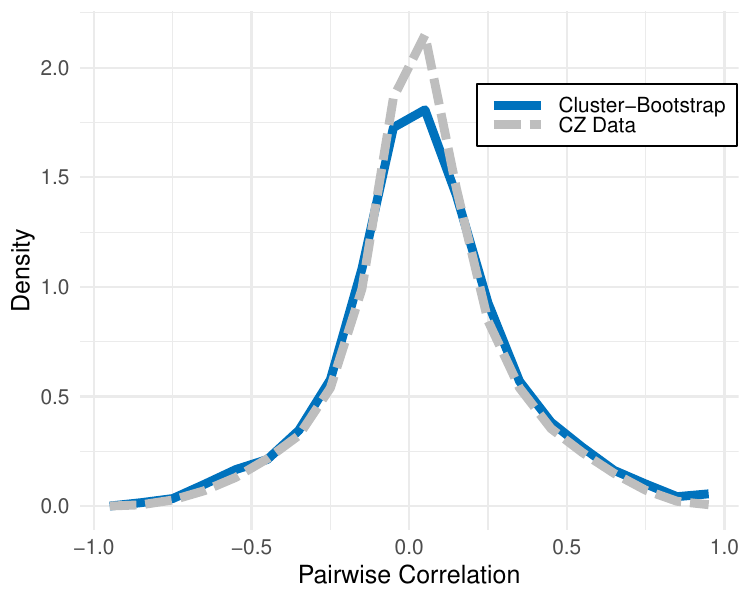} 
\end{figure}

I draw residuals with a cluster bootstrap. I draw 5,000 predictors
(with replacement) and 350 months (with replacement) and construct
a dataset of the corresponding residuals. 350 months corresponds to
the average sample size in the original papers. By construction, the
distribution of correlations in the bootstrapped samples is similar
that of the original data, as seen in Figure \ref{fig:semipar-cor}.

I then construct sampling noise $\varepsilon_{i}$ from these residuals
by taking the mean, dividing by the standard deviation, and multiplying
by the square root of the number of observations. I feed these $\varepsilon_{i}$
into the point estimate. That is, I simulate 5,000 $\mu_{i}$ using
the point estimate, calculate $t_{i}=\mu_{i}+\varepsilon_{i}$, and
then obtain published $t_{i}$ by applying Equation (\ref{eq:emp-pub-prob}),
once again using the point estimate. Provided that the model is well-specified,
this semi-parametric bootstrap provides valid inference (\citet{efron1994introduction},
Chapter 6.5).

This relatively complicated bootstrap is required for capturing correlations
because a simple cluster bootstrap would lead to t-statistics that
are too small compared to the empirical data, as it fails to simulate
publication bias. Similarly, a cluster bootstrap that simply applies
a truncation would lead to t-statistics that are too large compared
to empirical data, as this bootstrap effectively assumes that no publication
bias in the empirical data.

\pagebreak{}

\pdfbookmark{References}{ref}\bibliographystyle{elsarticle-harv}
\bibliography{0C__Users_ayc16_Dropbox_t-hurdles_2024-04-MS-Accept_lyx-thurdles-2024-04_chenbib}

\pagebreak{}

\setcounter{table}{0} \renewcommand{\thetable}{\arabic{table}}
\setcounter{figure}{1} \renewcommand{\thefigure}{\arabic{figure}}

\begin{figure}[h!]
\caption{\textbf{Weak Identification of t-Hurdles in Harvey, Liu, and Zhu (2016).
}HLZ baseline is the baseline estimate in Harvey, Liu, and Zhu's Table
5 (2nd row), which implies that t-hurdles should be raised to 2.27
to ensure FDR $\le$ 5\%. The alternative model uses the same model
but changes the share of false factors ($\pi_{F}$) to zero, implying
that even a t-hurdle of 0 ensures FDR $\le5\%$. HLZ assume only t-stats
> 2.57 are well-observed and both models are re-scaled to have the
same density in this region. Despite having very different implications
for t-hurdles, the two models are essentially observationally equivalent,
consistent with Proposition \ref{prop:weak}.}
\label{fig:hlz} \pdfbookmark{Figure 2}{hlz demo}\vspace{0.15in}

\centering
\includegraphics[width=0.9\textwidth]{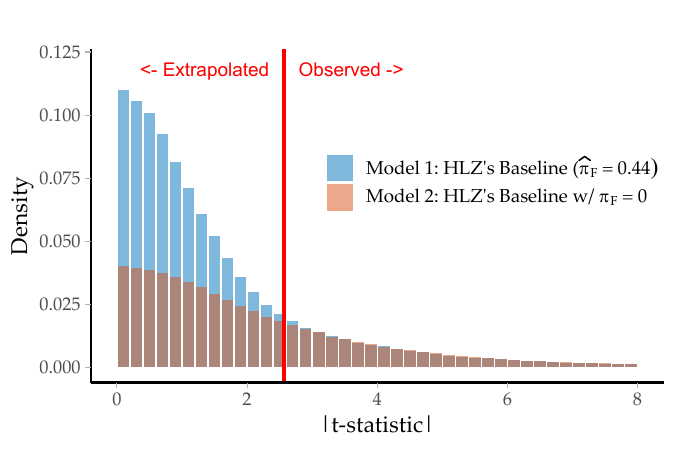}      
\end{figure}

\clearpage{}

\begin{figure}[h!]
\caption{\textbf{Model Fit and Intuition. }I examine predictions of the point
estimate (blue) and an estimate that assumes $\pi_{F}=2/3$ (red).
$\pi_{F}$ is the fraction of all predictors that are false. Panel
(a) compares model predictions for all t-stats (lines) with published
t-stats in the CZ data (gray bars). Panel (b) decomposes model predictions
into true predictors (bars) and false predictors (the space between
lines and bars) $\pi_{F}=0.01$ and $\pi_{F}=2/3$ both fit the data
very well, consistent with Proposition \ref{prop:weak}. Consistent
with Proposition \ref{prop:strong}, $\pi_{F}=0.01$ and $\pi_{F}=2/3$
imply similar distributions for true predictors, suggesting $\lambda$
is strongly identified.}
\label{fig:fit} \pdfbookmark{Figure 3}{point est} \vspace{0.15in}

\centering
\subfloat[Model Predictions vs Data]{\includegraphics[width=0.95\textwidth]{
	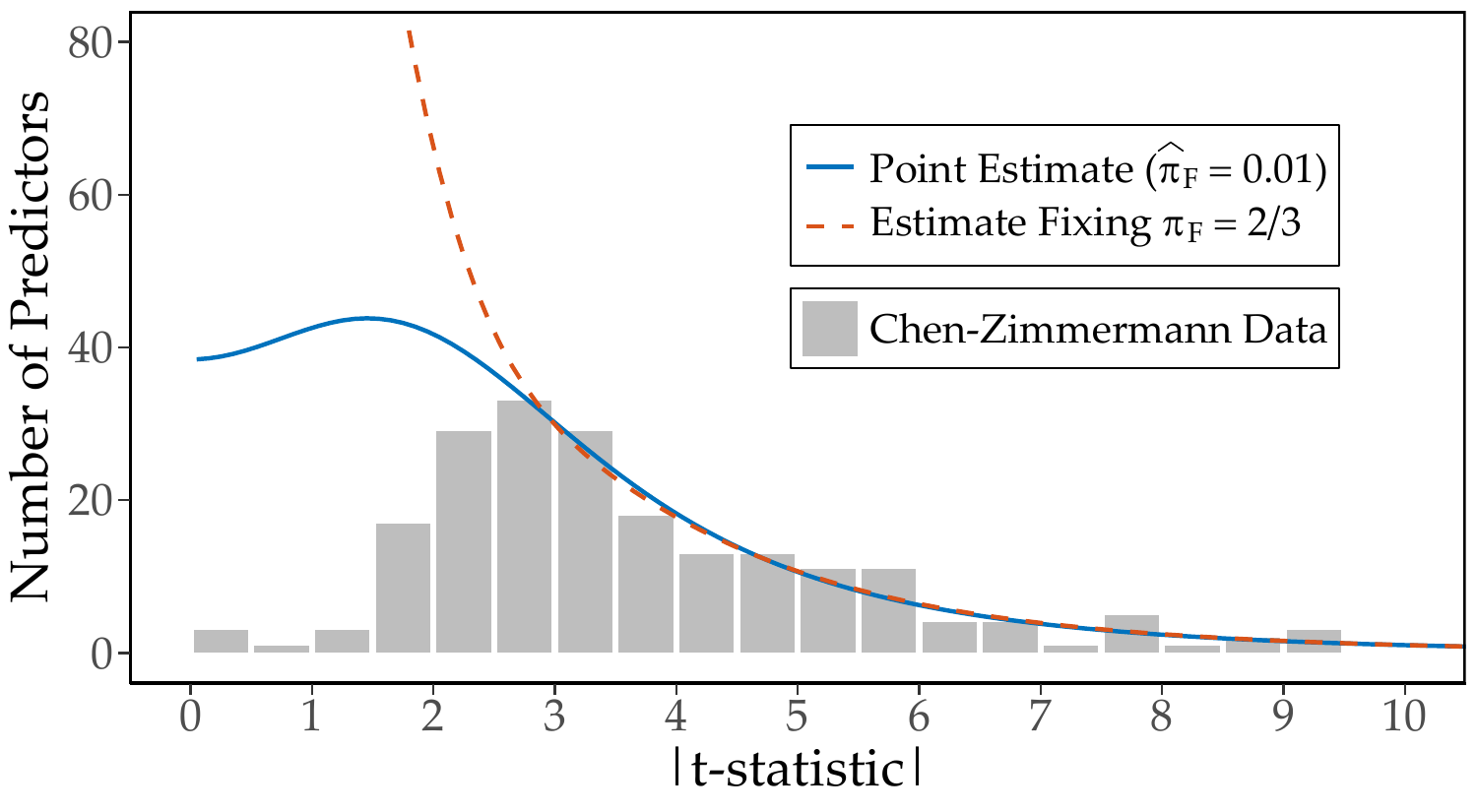
}} \\
\subfloat[Decomposition of Model Predictions]{\includegraphics[width=0.95\textwidth]{
  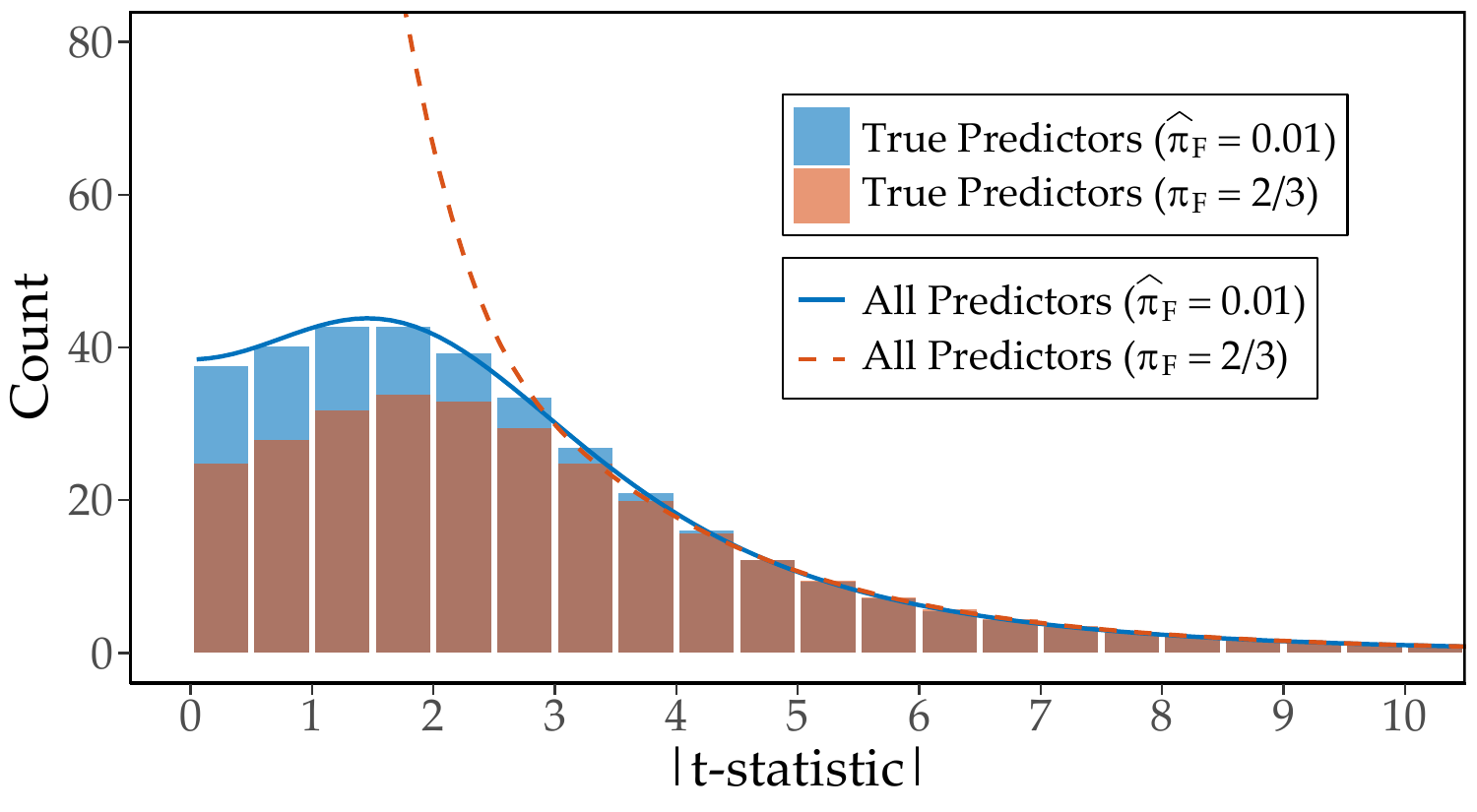
}}
\end{figure}

\clearpage{}
\begin{figure}[h!]
\caption{\textbf{Weak Identification of t-hurdles for Cross-Sectional Predictors.}
I repeatedly re-sample from the CZ dataset, re-estimate the model
of multiple testing with publication bias, and re-calculate the t-hurdle
(Equation (\ref{eq:theory-hurdle})) corresponding to an FDR of 5\%
(Panel (a)) and 1\% (Panel (b)). $\hat{\pi}_{F}$ is the estimated
probability of drawing a false factor. In both panels, the distribution
of bootstrapped multiple-testing-adjusted t-hurdles is highly dispersed,
with significant mass on both sides of the classical hurdles. Identification
is so weak that the data say little about whether t-stat hurdles should
be raised, stay the same, or even be lowered.}
\label{fig:boot-hurdles} \pdfbookmark{Figure 4}{boot hurdles}\vspace{0.15in}

\centering
\subfloat[FDR $\le$ 5\%]{\includegraphics[width=0.95\textwidth]{
  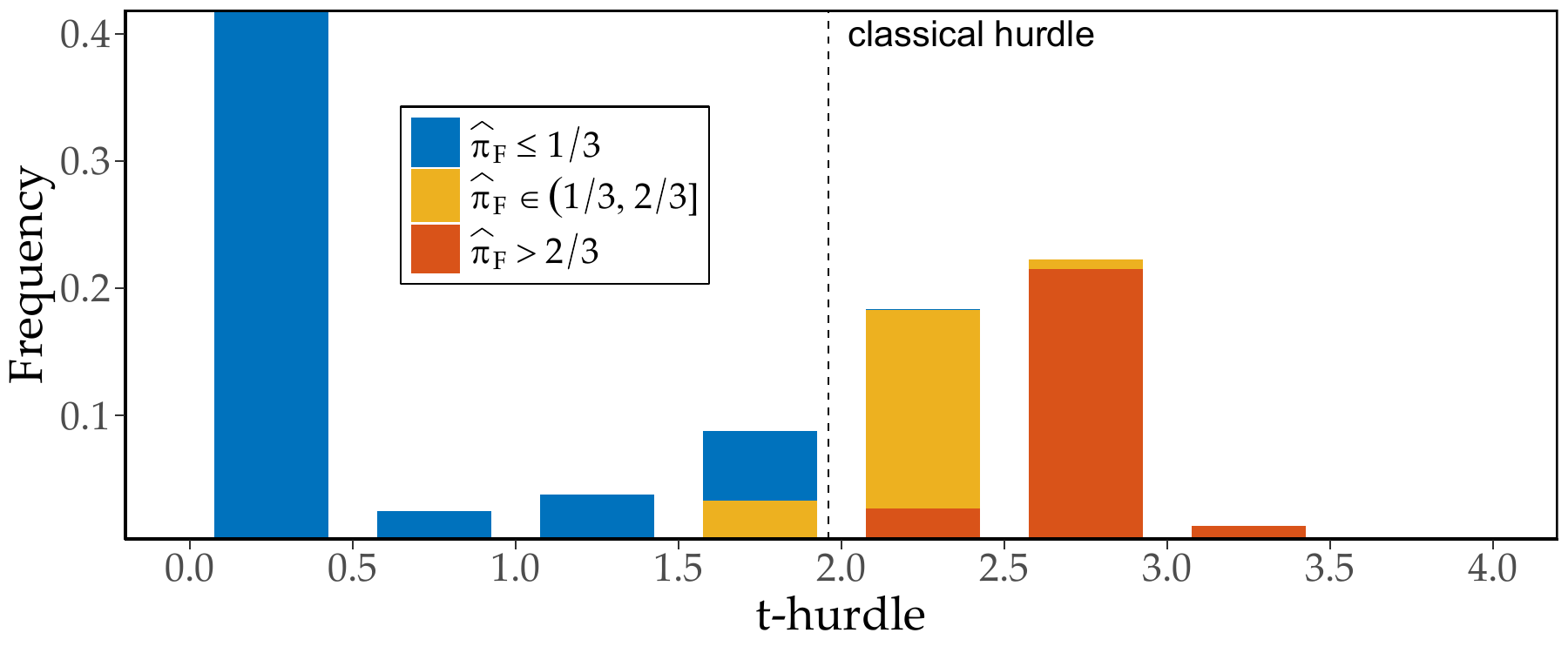
}}       \\
\subfloat[FDR $\le$ 1\%]{\includegraphics[width=0.95\textwidth]{
  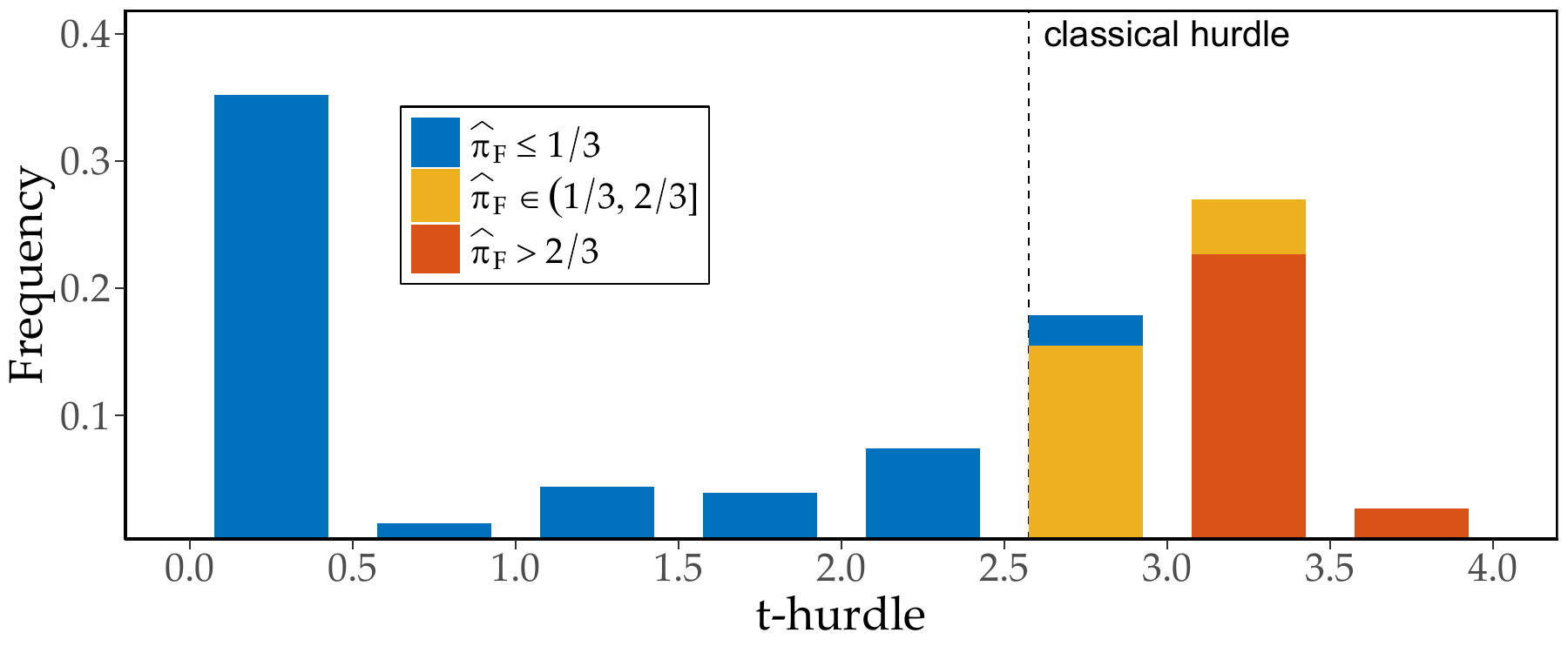
}}
\end{figure}

\clearpage{}
\begin{figure}[h!]
\caption{\textbf{Strong Identification of the Shrinkage and the local FDR.}
I repeatedly re-sample from the CZ dataset, re-estimate the model
of multiple testing with publication bias, and re-calculate shrinkage
(Equation (\ref{eq:theory-shrink-def}), Panel (a)) and the local
FDR (Equation (\ref{eq:theory-localfdr}), Panel (b)). Shrinkage is
the extent to which t-stats are biased upward due to multiple testing.
The local FDR is the probability a given predictor has an expected
return of zero. Within each bootstrap, shrinkage and local FDR are
averaged across published predictors. The McLean and Pontiff (2016)
bound is taken from their abstract. The HLZ Baseline applies Equation
(\ref{eq:theory-localfdr}) to simulated data based on their Table
5. Both multiple testing statistics imply that published predictability
is largely true with high confidence and are consistent with external
point estimates.}
\label{fig:boot-bias.pub-fdr} \pdfbookmark{Figure 5}{pub stats}\vspace{0.15in}

\centering
\subfloat[Average Shrinkage for Published t-Stats]{\includegraphics[width=0.95\textwidth]{
  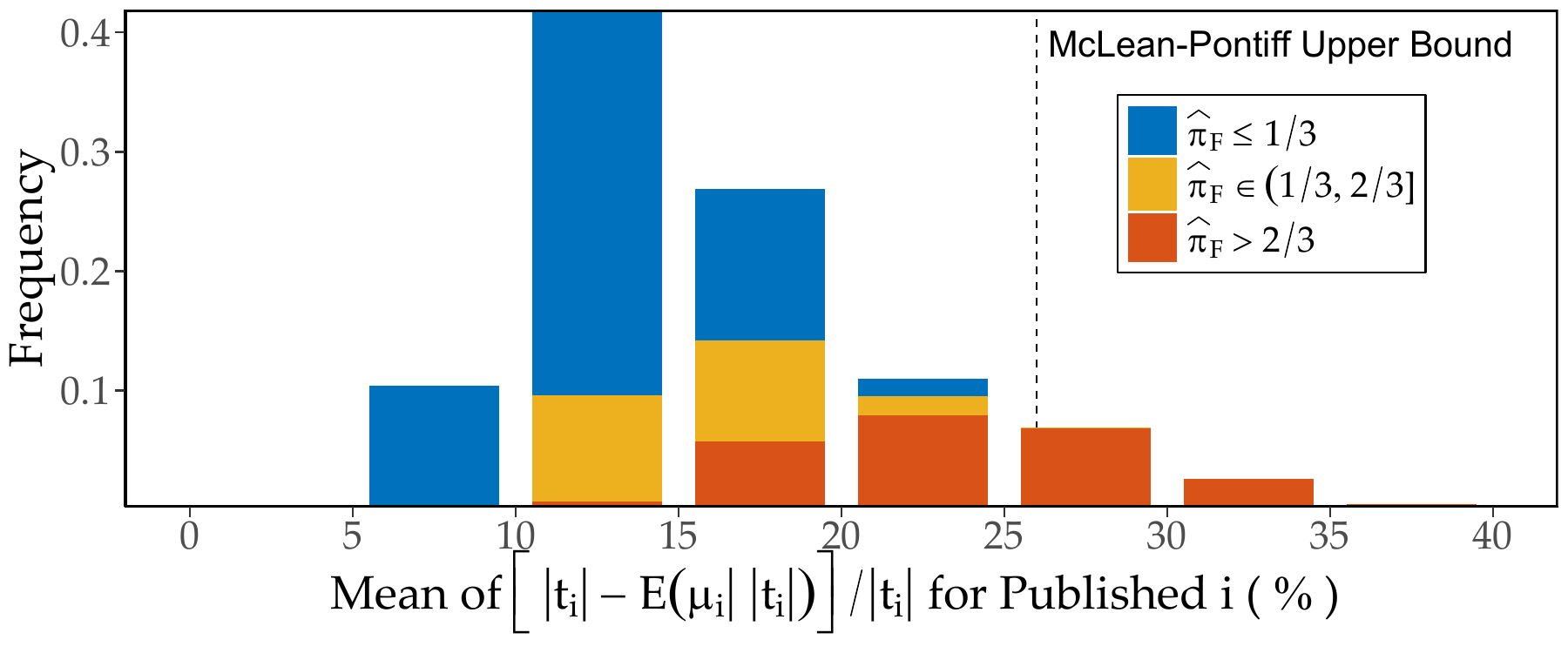
}}       \\
\subfloat[Local FDR for Published Predictors]{\includegraphics[width=0.95\textwidth]{
  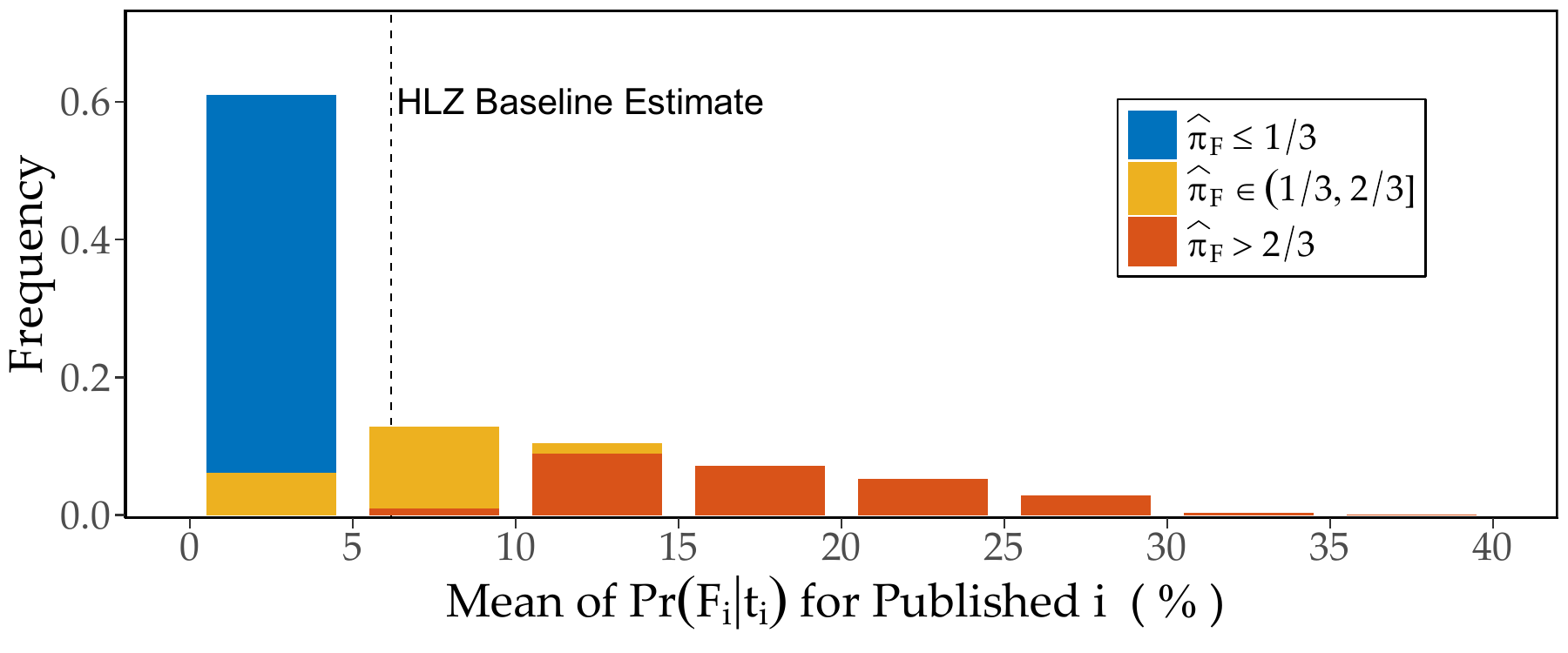
}}   
\end{figure}

\clearpage{}

\begin{table}[h!]
\caption{\textbf{Estimates of a Structural Model of the Cross-Sectional Literature}}
\label{tab:estimates} \pdfbookmark{Table 1}{parameter estimates}\vspace{0.1in}

{\small{}I estimate a structural model of multiple testing with publication
bias using quasi-maximum-likelihood (Section \ref{sec:emp-model})
on long-short portfolios from \citet*{ChenZimmermann2021} (Section
\ref{sec:emp_data}). The bootstrapped distribution is found by either
re-sampling the empirical t-stats and re-estimating (no parentheses)
or using a semi-parametric cluster-bootstrap that accounts for correlations
(parentheses). $\E\left(\mu_{i}|T_{i}\right)$ and $\SD\left(\mu_{i}|T_{i}\right)$
use textbook lognormal formulas to convert $\lambda_{\mu}$ and $\lambda_{\sigma}$
to moments. The share of false factors $\pi_{F}$ is weakly identified,
with huge confidence bounds, consistent with Proposition \ref{prop:weak}.
In contrast, $\E\left(\mu_{i}|T_{i}\right)$ and $\SD\left(\mu_{i}|T_{i}\right)$
are strongly identified. Accounting for correlations has very little
effect on the results.}{\small\par}
\begin{centering}
\vspace{0ex}
\par\end{centering}
\centering{}\setlength{\tabcolsep}{0.8ex} 
\begin{center} 
% ==== paste below

% Table generated by Excel2LaTeX from sheet 'estimates' 
\begin{tabular}{ccccccccccc} \toprule   &   & Point &   & \multicolumn{7}{c}{Bootstrapped Distribution} \\ \cmidrule{5-11}  &   & Estimate &   & 5 &   & 25 & 50 & 75 &   & 95 \\ \cmidrule{1-3}\cmidrule{5-5}\cmidrule{7-9}\cmidrule{11-11}\multicolumn{2}{l}{Probability false $\pi_F$} & 0.01 &   & 0.01  &   & 0.01  & 0.22  & 0.67  &   & 0.89  \\   &   &   &   & (0.01) &   & (0.01) & (0.31) & (0.65) &   & (0.85) \\ \multicolumn{11}{l}{True factors' unbiased t-stats ($\mu_i|T_i$)} \\   & \multicolumn{1}{l}{$\E(\mu_i|T_i)$} & 2.66 &   & 2.00  &   & 2.47  & 2.79  & 3.21  &   & 3.82  \\   &   &   &   & (2.00) &   & (2.46) & (2.80) & (3.13) &   & (3.56) \\   & \multicolumn{1}{l}{$\text{SD}(\mu_i|T_i)$} & 2.29 &   & 1.88  &   & 2.09  & 2.25  & 2.43  &   & 2.70  \\   &   &   &   & (1.89) &   & (2.13) & (2.26) & (2.44) &   & (2.67) \\   &   &   &   &   &   &   &   &   &   &  \\ \multicolumn{2}{l}{$\Pr(\pub_i)$ if marginal ($\eta$)} & 0.67 &   & 0.33  &   & 0.46  & 0.63  & 0.67  &   & 0.67  \\   &   &   &   & (0.33) &   & (0.46) & (0.60) & (0.67) &   & (0.67) \\ \bottomrule \end{tabular}% 

% ==== end paste
\end{center} 
\end{table}

\clearpage{}

\begin{table}[h!]
\caption{\textbf{Robustness}}
\label{tab:robust} \pdfbookmark{Table 2}{robustness}\vspace{0.1in}

{\small{}Each row shows the result of bootstrapping t-stat hurdles
}(Equation (\ref{eq:theory-hurdle})){\small{}, mean shrinkage for
published t-stats} (Equation (\ref{eq:theory-shrink-def}){\small{}),
and mean local FDR for published t-stats }(Equation (\ref{eq:theory-localfdr})){\small{}
using a different set of modeling assumptions and/or data-inclusion
requirements. Baseline uses the assumptions found throughout Section
\ref{sec:emp}. ``P05'' and ``P95'' show 5th and 95th percentiles
across 500 bootstrapped estimations for the robustness tests and 1000
bootstrapped estimations for the baseline. Across 10 alternative modeling
assumptions, the t-hurdle is weakly identified while the shrinkage
and local FDR are strongly identified.}{\small\par}
\begin{centering}
\vspace{0ex}
\par\end{centering}
\centering{}\setlength{\tabcolsep}{1.0ex} \small
\begin{center} 
% ==== paste below

% Table generated by Excel2LaTeX from sheet 'robustness' 
\begin{tabular}{clcccccccc} \toprule   &   & \multicolumn{2}{c}{t-hurdle} &   & \multicolumn{2}{c}{Shrinkage} &   & \multicolumn{2}{c}{FDR} \\   &   & \multicolumn{2}{c}{FDR = 5\%} &   & \multicolumn{2}{c}{for Pubs} &   & \multicolumn{2}{c}{for Pubs} \\ \cmidrule{3-4}\cmidrule{6-7}\cmidrule{9-10}  &   & P05 & P95 &   & P05 & P95 &   & P05 & P95 \\ \cmidrule{3-4}\cmidrule{6-7}\cmidrule{9-10}(1) & Baseline & 0.0 & 3.0 &   & 8.9 & 27.9 &   & 0.1 & 22.6 \\   &   &   &   &   &   &   &   &   &  \\ \multicolumn{10}{l}{Including $|t_i| <1.96$} \\ (2) & 3-step $\Pr(\pub_i)$ & 0.0 & 1.6 &   & 2.6 & 17.7 &   & 0.1 & 4.6 \\ (3) & Restricted 3-step $\Pr(\pub_i)$ & 0.0 & 2.4 &   & 7.8 & 22.6 &   & 0.1 & 15.7 \\   &   &   &   &   &   &   &   &   &  \\ \multicolumn{10}{l}{Excluding the $\pi_F$ very small} \\ (4) & $\pi_F \ge 0.1$ & 0.9 & 3.0 &   & 9.0 & 27.8 &   & 0.6 & 23.3 \\ (5) & $\pi_F \ge 0.2$ & 1.5 & 3.0 &   & 9.2 & 26.6 &   & 1.4 & 22.0 \\   &   &   &   &   &   &   &   &   &  \\ \multicolumn{10}{l}{Alternative publication probability function} \\ (6) & $\eta = 0.5$ & 0.0 & 2.9 &   & 10.9 & 26.1 &   & 0.1 & 20.7 \\ (7) & $\eta \in [1/3, 1]$ & 0.0 & 3.0 &   & 6.9 & 26.8 &   & 0.1 & 22.6 \\ (8) & logistic $\Pr(\pub_i)$ & 0.0 & 2.7 &   & 6.6 & 21.2 &   & 0.1 & 17.1 \\   &   &   &   &   &   &   &   &   &  \\ \multicolumn{10}{l}{Alternative distributional assumptions} \\ (9) & $\mu_i|T_i \sim \text{Exponential}$ & 0.0 & 2.9 &   & 11.3 & 28.7 &   & 0.1 & 19.7 \\ (10) & $\mu_i|T_i \sim \text{Student's t}$ & 0.0 & 2.9 &   & 8.1 & 22.9 &   & 0.1 & 20.9 \\ (11) & $\mu_i|T_i \sim \text{Mixture-Normal}$ & 0.0 & 3.0 &   & 7.2 & 24.1 &   & 0.1 & 24.8 \\ \bottomrule \end{tabular}% 

% ==== end paste
\end{center} 
\end{table}

\end{document}